\documentclass[aps,pra,nolongbibliography,notitlepage,twocolumn,superscriptaddress]{revtex4-2}
\usepackage{siunitx}
\usepackage{amsmath}
\usepackage{amssymb}
\usepackage{bbold}
\usepackage{color}
\usepackage{comment}
\usepackage{tikz}
\usepackage{pgfplots}
\pgfplotsset{compat=1.5}
\usepackage{graphicx}
\usepackage[colorlinks=true, linkcolor=black, citecolor=blue, pdfencoding=auto]{hyperref}
\usepackage{blindtext}
\usepackage{physics}

\newcommand{\bk}{\bf k}
\newcommand{\bq}{\bf q}
\def\phdag{{\phantom \dagger}}

\newcommand{\rev}[1]{ {\color{black} #1}}

\begin{document}
\title{Geometric Stiffness in Interlayer Exciton Condensates}

 \author{Nishchhal Verma}
 \thanks{These authors contributed equally}
\affiliation{Department of Physics,
Columbia University,
New York, NY 10027, USA}
\affiliation{Center for Computational Quantum Physics, Flatiron Institute, New York, New York 10010, USA}
\author{Daniele Guerci}
\thanks{These authors contributed equally}
\affiliation{Center for Computational Quantum Physics, Flatiron Institute, New York, New York 10010, USA}
\author{Raquel Queiroz}
\email{raquel.queiroz@columbia.edu}
\affiliation{Department of Physics,
Columbia University,
New York, NY 10027, USA}
\affiliation{Center for Computational Quantum Physics, Flatiron Institute, New York, New York 10010, USA}

\begin{abstract}
Recent experiments have confirmed the presence of interlayer excitons in the ground state of transition metal dichalcogenide (TMD) bilayers. 
The interlayer excitons are expected to show remarkable transport properties when they undergo Bose condensation.
In this work, we demonstrate that quantum geometry of Bloch wavefunctions plays an important role in the phase stiffness of the Interlayer Exciton Condensate (IEC).
Notably, we identify a geometric contribution that amplifies the stiffness, leading to the formation of a robust condensate with an increased BKT temperature.
Our results have direct implications for the ongoing experimental efforts on interlayer excitons in materials that have non-trivial quantum geometry. 
We provide quantitative estimates for the geometric contribution in TMD bilayers through a realistic continuum model with gated Coulomb interaction, and find that the substantially increased stiffness allows for an IEC to be realized at amenable experimental conditions.
\end{abstract}

\maketitle

\paragraph*{Introduction.---}
Advancements in topological quantum matter have drawn attention to the crucial role of Bloch wavefunctions in diverse condensed matter systems. 
While the influence of Berry curvature on non-interacting electrons is well-understood as an anomalous velocity \cite{Xiao2010}, a closely related quantity, the quantum metric, has recently gained significant attention particularly in the context of flat-band superconductivity and related experiments in moiré heterostructures \cite{Torma2021}.
Derived from the geometric properties of Bloch wavefunctions, the quantum metric has profound effects on various facets of superconductivity.
Notably, it modifies the mass of Cooper pairs \cite{Torma2018,Iskin2018,Iskin2021}, phase stiffness \cite{Peotta2015,Julku2016,Liang2017,Xie2020, HerzogArbeitman2022,Huhtinen2022,Mao2023}, spectral weight \cite{Tovmasyan2016,Verma2021,Ahn2021,mao2023upper}, and potentially the critical temperature \cite{Hofmann2020,Peri2021}.
Other than superconductivity, quantum geometry is known to appear in current noise spectrum \cite{Neupert2013}, dielectric response \cite{komissarov2023}, electron-phonon coupling \cite{yu2023}, plasmons \cite{Arora2022} and nonlinear response \cite{Lapa2019,Gao2019,Kozii2021,Ahn2022,Chaudhary2022}.
In this work, we target Interlayer Exciton Condensates (IECs) and reveal a significant geometric contribution to the phase stiffness that results in a more robust condensate characterized by a higher Berezinskii-Kosterlitz-Thouless (BKT) transition temperature.

Excitons, bound states of electrons and holes in semiconductors, are bosons that have long been proposed to form a Bose-Einstein condensate (BEC) at low temperatures \cite{Blatt1962,Halperin1968}. 
Unlike conventional BECs, exciton condensates conserve total particle number and instead break a $U(1)_e \times U(1)_h$ symmetry that corresponds to separate conservation of electrons and holes \cite{lozovik1975feasibility, lozovik1976new}.
This symmetry is experimentally realizable in a bilayer system with a spacer that suppresses single-particle tunneling between the layers.
If the electrons reside on the top $(t)$ and holes in bottom $(b)$ layers, IEC is formed by the spontaneous breaking of $U(1)_t \times U(1)_b$ symmetry \cite{Yoshioka1990,Yang2001,De2002,Joglekar2006}.
Its superfluid properties have been observed in quantum Hall systems~\cite{Eisenstein2004,Eisenstein2014}.

Although an exciton condensate arising intrinsically in a real material has been a challenge, there has been progress in three-dimensional semimetal 1T-${\rm TiSe}_2$~\cite{Kogar2017}, \rev{monolayer ${\rm WTe}_2$ \cite{Jia2022,Sun2022}, bilayer ${\rm WSe}_2 $ \cite{Shi_2022}}, and TMD bilayer ${\rm WSe}_2/{\rm MoSe}_2$ \cite{Wang2019}. 
In particular, ref.~\cite{Ma2021} has established the existence of interlayer excitons in the ground state by capacitance measurements and characterized the exciton Mott transition~\cite{mott1949basis, Mott1973, Brinkman_1973, Smith1986, Guerci2019} as a function of the density of electron-hole pairs.
These interlayer excitons have finite dipole moment and interact via dipole-dipole interaction.
Thus, as a virtue of interacting bosons in 2D, it is possible that there is condensation at low enough temperatures~\cite{nozieres1985bose, Fogler2014, Wu2015, Berman2016, Debnath2017}.
Confirming the existence of the condensate requires transport experiments \cite{lozovik1976new,High2012}, which have remained elusive \rev{until two recent reports \cite{Nguyen2023,Qi2023}}.

IECs display fascinating transport properties, most notably dissipationless counterflow transport.
When equal and opposite fields are applied to the two layers, excitons flow without resistance in the condensate \cite{Eisenstein2004}. 
The longitudinal counterflow conductivity, $\sigma_{\rm CF}(\omega)$, diverges in the dc limit with $\sigma_{\rm CF}(\omega) = D_s \delta(\omega) + \cdots$, where the weight of the delta function, $D_s$, represents the phase stiffness that governs the free energy cost of phase fluctuations in the condensate \cite{scalapino1993}.
In this study, we demonstrate that $D_s$, in addition to a conventional contribution from band dispersion, has a significant geometric contribution that arises from the wavefunctions of the non-interacting electron and hole bands.

The critical role of wavefunctions in exciton condensation is evident from quantum Hall bilayers where excitons exhibit macroscopic coherence and condense \cite{Fertig1989,Spielman2001} despite the non-interacting bands having flat dispersion and infinite mass.
The non-trivial wavefunctions of the Landau level endow mobility to excitons even when the constituent electrons and holes are immobile.
Similar phenomenology is anticipated in materials without a magnetic field, such as ${\rm WSe}_2/{\rm MoSe}_2$, as a virtue of their non-trivial wavefunctions \cite{Xiao2007,Cao2012,Wang2018}.
However, there are key distinctions due to finite dispersion and asymmetric electron-hole bands, and it remains to be seen if such effects preserve or diminish the geometric contribution.
We address these fundamental questions in this paper.

We begin by deriving the phase stiffness of a general exciton Hamiltonian, separating it into conventional and geometric components.
The latter arises primarily from the non-interacting electron/hole wavefunctions.
\rev{To demonstrate this effect, we consider a realistic continuum model endowed with screened Coulomb interactions applicable to bilayer TMD devices and provide estimates for the geometric contribution to stiffness} in ${\rm MoTe}_2$ homobilayers \cite{Robert2016} as well as ${\rm WSe}_2/{\rm MoSe}_2$ heterobilayers \cite{Ma2021}. 
Finally, we propose experimental setups that can serve as validation platforms for our theory.

\paragraph*{Phase Stiffness.---}
Thermodynamic stability of a condensate depends on the free energy cost associated with spatial fluctuations in the phase, $\theta \rightarrow \theta({\bf r})$, which is quantified as $\mathcal{F} = (1/2) D_s \int d{\bf r}|\boldsymbol\nabla \theta({\bf r})|^2$, where $D_s$ represents the phase stiffness. 
For IECs, the phase stiffness is calculated as a linear response coefficient when equal and opposite vector potentials are applied to the two layers \cite{Eisenstein2014}.
The antisymmetric vector potential couples symmetrically to the exciton, since electron and holes have opposite charges, inducing infinitesimal phase fluctuations in the condensate.

\begin{figure}
    \centering
    \includegraphics[width=0.4\textwidth]{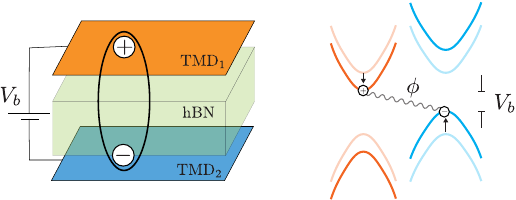}
    \caption{ 
     Device setup with the two layers, orange and blue, with electrons in one and holes in the other. The hBN spacer (green) suppresses single-particle tunneling between the layers. Bias Voltage, $V_b$, tunes the gap between lowest unoccupied band and highest occupied band. It is important that these are direct excitons. The bands are shifted horizontally to highlight the layers.}
    \label{fig:idea}
\end{figure}

A general model Hamiltonian for the IEC is given by $\mathcal{H}_{\rm ex} = \mathcal{H}_0 + \mathcal{H}_{\rm int}$, where $\mathcal{H}_0 = \mathcal{H}_t + \mathcal{H}_b$ describes the non-interacting properties with Bloch Hamiltonians $\{\mathcal{H}_{\nu}\}$ for the two layers $\nu = \{t,b\}$, and $\mathcal{H}_{\rm int}$ is a density-density interaction that gives rise to excitons.
We write $\mathcal{H}_0 = \sum_{\bf k} \Psi^\dag_{{\bf k}} H_0({\bf k}) \Psi^\dag_{{\bf k}}$ where $H_0({\bf k}) = [H_{t}({\bf k}) \oplus H_{b}({\bf k})] + V_b \tau_z$ and $\Psi^{\phantom\dag}_{{\bf k}}$ is a spinor that has internal labels $(\alpha,\beta)$ that are omitted for brevity along with layer label $\nu$: $\Psi^{\phantom\dag}_{{\bf k}} = (c^\phdag_{ {\bf k}, \alpha, t }, \cdots,  c^\phdag_{ {\bf k}, \beta, b }, \cdots )^T$, and ${\bf k} \in {\rm BZ}$.
The last term describes a bias voltage $V_b$ that tunes the gap between the conduction and valence bands (see Fig.~\ref{fig:idea}) \rev{with $\tau_z$ being the Pauli matrix in the layer subspace}.
With vanishing single-particle tunneling between the layers, the non-interacting model has a $U(1)_t \times U(1)_b$ symmetry that is spontaneously broken by the IEC.

Density-density interactions remain unaffected by the external vector potential. 
Consequently, when equal and opposite ${\bf A} = A\hat{x}$ are applied to the two layers, 
the current operators depend solely on the Bloch Hamiltonian as $H_0({\bf k}, A) = H_t({\bf k}-eA/\hbar) \oplus H_b({\bf k}+eA/\hbar)$.
We assume spatially isotropic systems to suppress the tensor nature of currents and stiffness and leave the extension to anisotropic systems to App.~A6.
The current operator can be expressed as $j = -\delta\mathcal{H}_0/\delta A = j_P + A j_D$, where $j_P = (e/\hbar) \sum_{ {\bf k} } \Psi^\dag_{ {\bf k} } \left[ \tau^z \partial_{k}\mathcal{H}_{0}({\bf k}) \right] \Psi^\phdag_{{\bf k}}$ and $j_D = -(e/\hbar)^2 \sum_{ {\bf k}} \Psi^\dag_{ {\bf k}} \partial_{k}^2 \mathcal{H}_{0}({\bf k}) \Psi^\phdag_{{\bf k}}$ are the paramagnetic and diamagnetic currents respectively.
The stiffness is determined by the Kubo formula, given by $D_s = -[ \langle j_D \rangle - \chi_{ j_P j_P}({\bf q}_\perp\rightarrow 0, \omega=0)]/4$, where $\chi_{ j_P j_P}$ is the longitudinal current-current correlator \cite{scalapino1993,Benfatto2004,Cea2015}. 
Calculating the stiffness requires a complete enumeration of the eigenstates of the full interacting Hamiltonian $\mathcal{H}_{\rm ex}$.
To make progress beyond the formal definition, we focus on mean-field theories where the interaction term breaks down into fermionic bilinears. 
We can then utilize the eigenspectrum $E_{m,{\bf k}}$ and eigenstates $|u_{m,{\bf k}}\rangle$ of the mean-field Hamiltonian to express the stiffness as:
\begin{multline}
    D_s = \dfrac{\hbar^2}{4e^2 A} \Big[ \sum\limits_{ {\bf k}, m} f_{m,{\bf k}} \left\langle u_{m, {\bf k}} \left|  \partial^2_k H_0({\bf k}) \right| u_{m, {\bf k}} \right\rangle - \\
    \sum\limits_{{\bf k}, m, n} \dfrac{f_{m,{\bf k}}-f_{n,{\bf k}}}{E_{n,{\bf k}}-E_{m,{\bf k}}}  \left|\left\langle u_{m, {\bf k}} \left| \tau^z \partial_k H_0({\bf k})\right| u_{n, {\bf k}} \right\rangle\right|^2\Big] \label{eq:Ds-Multi-Band}
\end{multline}
where $f_{m,{\bf k}}\equiv f[E_{m,{\bf k}}]$ is the Fermi occupation factor and $A$ is the volume normalization factor which is the area of the sample in 2D.
\rev{The stiffness measures coherence between transport in the two layers.
In the absence of inter-layer tunneling, the only way to get finite stiffness is for the interaction to admit an off-diagonal exciton term $ \Psi^\dag_{{\bf k}} [\tau^i \hat{\phi}({\bf k})] \Psi^\phdag_{{\bf k}}$ with $i= x, y$ within mean-field. 
The matrix function $\hat{\phi}({\bf k})$ is a mean-field ansatz that needs to be evaluated self-consistently.}

Although the dependence of the phase stiffness on wavefunctions is evident in Eq.~\eqref{eq:Ds-Multi-Band} from the matrix structure of the current operators, the geometric and energetic terms are intertwined with no clear route to separation.
To tackle this challenge, we introduce a projected low-energy model \cite{Xie2020}, which is applicable when excitons predominantly arise from the \rev{lowest lying electron band ($e$) and the highest hole band ($h$)} 
\begin{equation}
\mathcal{H}_{\rm ex} =  \sum\limits_{\bf k} \psi^\dag_{{\bf k}} \begin{pmatrix}
         \epsilon_e({\bf k}) - \Sigma_e({\bf k}) & \varphi({\bf k}) \\  \varphi({\bf k})^* &  \epsilon_h({\bf k}) + \Sigma_h({\bf k})
    \end{pmatrix} \psi^{\phantom\dag}_{{\bf k}} \label{eq:projMF}
\end{equation}
where $\epsilon_{e/h}$ are the bare electron/hole dispersions, $\psi^\phdag_{\bf k} = ( c^\phdag_{ {\bf k}, e}, c^\phdag_{ {\bf k}, h} )^T$ is the low-energy basis state,  $\Sigma_{e/h}({\bf k})$ are the self-energies (including $V_b$) and $\varphi({\bf k})$ is a shorthand for
\begin{equation}
    \varphi({\bf k}) = \sum\limits_{ \alpha\in t, \beta \in b} [U_t({\bf k})]^*_{e, \alpha } \phi_{\alpha\beta}({\bf k}) [U_b({\bf k})]_{\beta, h } .
\end{equation}
A crucial aspect to highlight is that $U_t({\bf k})$ and $U_b({\bf k})$ are independent unitary matrices that diagonalize $H_t({\bf k})$ and $H_b({\bf k})$ respectively.
The subscripts $e$ and $h$ indicate the electron and hole bands involved in the exciton pairing. 
It is important to note that, unlike in BCS theory where the redundancy in Nambu basis introduces a particle-hole symmetry in the BdG Hamiltonian, $U_t({\bf k})$ and $U_b({\bf k})$ do not have such constraints. 
Hence our framework is a generalization of phase stiffness, that is designed for ${\rm WSe}_2/{\rm MoSe}_2$ bilayers where the electron and hole bands originate from distinct materials.

We next assume $\epsilon_e({\bf k}) = -\epsilon_h({\bf k}) = \epsilon({\bf k})$, while allowing $U_t({\bf k})$ to differ from $U_b({\bf k})$. 
This assumption is justified as the effect of asymmetric dispersion on exciton is a well understood textbook problem \cite{Fox2001} that only complicates our analysis without offering additional insight  (see also App.~C).
On the other hand, the presence of $U_t({\bf k})$ and $U_b({\bf k})$ has a non-trivial consequence for $\varphi({\bf k})$ in the presence of vector potential:
\begin{equation}
\varphi({\bf k}, A) \approx \varphi({\bf k}) - \dfrac{eA}{\hbar} \mathcal{P}({\bf k}) - \dfrac{e^2 A^2}{2\hbar^2} \mathcal{D}({\bf k}),
\end{equation}
where $\mathcal{P}({\bf k})$ and $\mathcal{D}({\bf k})$ are the paramagnetic and diamagnetic terms that involve derivatives of $U_\nu({\bf k})$ (see App.~A2 for the full expression).
These terms are in addition to the energetic terms arising from $\partial_k\epsilon({\bf k})$ and $\partial_k^2\epsilon({\bf k})$ in the current operator and are key to our analysis.
After performing a lengthy but straightforward calculation that is detailed in App.~A2, we find two main contributions to the stiffness $D_s = D_s^c + D_s^g$ with
\begin{eqnarray}
    D_s^c &=& \dfrac{1}{2A}\sum\limits_{ {\bf k} } \partial_{k_i}^2 \epsilon({\bf k}) v_{\bf k}^2 \\ 
    D_s^g &=& \dfrac{1 }{4A}\sum\limits_{ {\bf k} } \dfrac{\mathcal{G}({\bf k})}{E({\bf k})}  + \dfrac{1 }{4A}\sum\limits_{ {\bf k} } \dfrac{{\rm Re}[\mathcal{P}({\bf k})\varphi({\bf k})^*]^2}{E({\bf k})^3}  \label{eq:Main-Result-1}
\end{eqnarray}
where $v_{{\bf k}}^2 = (1- \xi({\bf k})/E({\bf k}))/2$ is the momentum occupation factor with $\xi({\bf k}) = \epsilon({\bf k}) - \Sigma({\bf k})$, and $E({\bf k}) = \sqrt{\xi({\bf k})^2 + |\varphi({\bf k})|^2}$ is the mean-field quasiparticle dispersion. The second part of the equation includes the geometric quantity $\mathcal{G}({\bf k}) = {\rm Re}[\mathcal{D}({\bf k}) \varphi({\bf k})^*] -  \left| \mathcal{P}({\bf k})\right|^2 $.
We note that all terms in eq.~\eqref{eq:Main-Result-1} are individually gauge invariant (see App.~A4) and constitute the main result of our paper.

A few comments are in order.
The geometric contribution in Eq.~\eqref{eq:Main-Result-1} arises explicitly from the Taylor expansion of the off-diagonal term $\varphi({\bf k}, A)$. 
If the wavefunctions were trivial, say independent of ${\bf k}$, this contribution would vanish since $\mathcal{P}({\bf k}) = \mathcal{D}({\bf k}) = 0$.
\rev{
As one might expect, there are some connections with superconductivity as well. If the bands were particle-hole symmetric, eq.~\eqref{eq:projMF} would reduce to a BdG matrix and the function $\mathcal{G}({\bf k})$ will become precisely the quantum metric, $g({\bf k})$.
This is not surprising since superfluid stiffness is known to be enhanced by quantum metric in superconductors within mean-field \cite{Torma2021}.
Since the particle-hole symmetry is not enforced in an exciton mean-field theory, our results present a generalization of quantum geometric phase stiffness beyond pairing between particle-hole symmetric states.
}
We further note that orbital embedding modifies our calculation in a similar fashion as in superconductivity \cite{Huhtinen2022}. 
It enforces the minimal condition where the solution to gap equation is kept real in the presence of external vector potential \cite{Peotta2022}.
Our result demonstrate a generalization of excitonic phase stiffness beyond particle-hole symmetry discussed in the context of quantum Hall bilayers \cite{Fertig1989}.

\paragraph*{Hubbard Model with tunable quantum metric.---}
\rev{
To illustrate the phenomenology with a simple model, we consider an interlayer on-site orbital-diagonal Hubbard interaction  $\mathcal{H}_{\rm int} = V \sum_{{\bf i},\alpha} \hat{n}_{{\bf i}, \alpha, t} \hat{n}_{{\bf i}, \alpha, b}$.
The labels ${\bf i}, \alpha$ pertaining to unit cell index and orbitals, are not crucial for our discussion as long as the interaction is interlayer in character.
Details about the mean-field theory can be found in App.~B1.

For the Bloch hamiltonian,} we consider a tunable quantum metric model that has two orbitals on a square lattice with the Bloch Hamiltonian
\begin{equation}
    H(\zeta, {\bf k}) = 2t(2-p_{\bf k}) + t_F \cos(\zeta p_{\bf k}) \sigma^x + t_F \sin(\zeta p_{\bf k}) \sigma^y \label{eq:def:zeta-ham}
\end{equation}
where the Pauli matrices $\sigma^i$ act in orbital space, $p_{\bf k}$ is a periodic function given by $ p_{\bf k} = \cos k_x + \cos k_y$ and the parameter $\zeta$ controls the quantum geometry by introducing long range hoppings \cite{Hofmann2022}. 
More precisely, $\zeta$ is an overall scaling factor for the quantum geometric tensor $\mathcal{Q}( {\bf k}, \zeta )_{\mu\nu} = (\zeta^2/4) \sin k_\mu \sin k_\nu $, which is real because of inversion and time-reversal symmetries.
The Berry curvature $F({\bf k}) = -{\rm Im}[\mathcal{Q}( {\bf k})]_{xy}/2$ vanishes identically while the quantum metric $g_{\mu\nu}({\bf k}) = {\rm Re}[\mathcal{Q}_{ {\bf k} }]_{\mu\nu}$ is finite.
Another convenient aspect of this model is that the band dispersions $\epsilon_{\pm,{\bf k}} = 2t(2-p_{\bf k}) \pm t_F$ are independent of $\zeta$.
This permits the use of $\zeta$ to tune quantum geometry without affecting the band dispersion, capturing the discussion around Eq.~\eqref{eq:Main-Result-1}.

\begin{figure}
    \centering
    \includegraphics[width=0.45\textwidth]{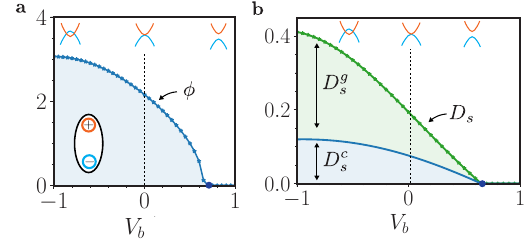}
    \caption{${\bf a}$ Mean-field gap solution for the toy model with tunable quantum metric at fixed interaction $U=6t$. ${\bf b}$. Mean-field Phase stiffness.  All energy scales, $\phi$, $D_s$, $V_b$ are plotted in units of single-particle hopping $t$ (see Eq.~\eqref{eq:def:zeta-ham}). Subscripts $c$ and $g$ denote conventional and geometric contributions respectively. }
    \label{fig:phi-Ds-minimal}
\end{figure}

Symmetric bilayers, where both layers are described by the same parameter $\zeta$, give rise to excitons (see Fig.\ref{fig:phi-Ds-minimal}{\bf a}) where the stiffness has a significant enhancement coming from the geometric term $D_s^g$. 
The extra contribution is indeed proportional to the trace of the quantum metric in this particular case, as
Eq.\eqref{eq:Main-Result-1} simplifies to $D_s^g = (\phi^2/4A) \sum_{\bf k} g({\bf k})/E({\bf k})$ \cite{Hu2022}.
While the geometric contribution dominates at higher bias voltage $|V_b|$, it is important to note that a high $|V_b|$ results in higher densities of electron-hole pairs and may ultimately drive the system to the exciton Mott transition \cite{Mott1973, Brinkman_1973, Smith1986, Guerci2019}.
The case of asymmetric wavefunctions follows along similar lines and is outlined in App.~C and Fig.~S3.

\paragraph*{Continuum Model for TMDs.---}
\rev{With the formalism in place, we now aim to quantify the role of quantum geometric effects in TMD exciton bilayers.
TMDs are semi-conductors with valley optical selection rules \cite{PhysRevLett.108.196802} and hence their minimal model is that of a gapped Dirac cone.
We consider the continuum Hamiltonian introduced in ref.~\cite{Rice_1995} with spin-valley locking.
The continuum $k.p$ Hamiltonian for a given spin and valley, to first order in $k$, reads
\begin{equation}
\label{kp_hamiltonian}
    H_{\nu}({\bk}) = \Delta_\nu\sigma_z + v_\nu\mathbf k\cdot\sigma,
\end{equation}
where $\sigma^i$ is the Pauli matrix in the internal sublattice degrees of freedom for the conduction and valence bands, and dispersion $\epsilon_{\bk\nu\pm}=\pm\sqrt{\Delta^2_\nu+ v_\nu^2 |{\bf k}|^2}$.
The two-band aspect of the model is justified for the valence band where the Ising spin-orbit coupling (SOC) gap is of the order of $ 180$meV in MoX$_2$ and $ 440$meV in WX$_2$~\cite{Korm_nyos_2015}. 
We also neglect the quadratic trigonal warping term and the quadratic particle-hole mass imbalance since it does not play an important role in our phenomenological discussion.}

The geometric properties of the model are encoded in the form factors $F^\nu_{{\bk},{\bq}}=U^\dagger_\nu({\bk})U_\nu({\bk}+{\bq})$ which has non-trivial momentum dependence.
\rev{In order to isolate the geometric contribution, we will compare our results with the gapped Dirac cone in eq.~\eqref{kp_hamiltonian} to that of a parabolic band $H_\nu (\bk) = \pm \epsilon_{\bk\nu} \sigma_z$ with energies $\epsilon_{\bk\nu}\approx \Delta_\nu + k^2/2m_\nu$ with $m_\nu=v^2_\nu/m_\nu$ and trivial form factor $F^\nu_{{\bk},{\bq}} = \delta_{{\bk},{\bq}} $.}

\begin{figure}
    \centering
    \includegraphics[width=0.47\textwidth]{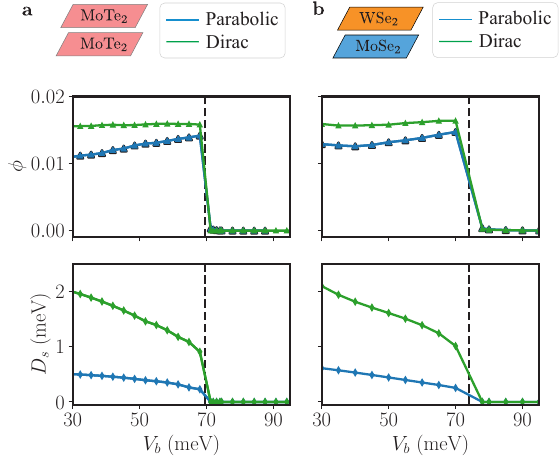}
    \caption{We show the interlayer exciton coherence $\phi$ and the phase stiffness $D_s$ as a function of $V_b$. Green and Blue data corresponds to a gapped Dirac cone and parabolic bands with mass $m=v^2/\Delta$, respectively. ${\bf a}$: Homobilayer with velocity $v=2.16$eV\AA, energy gap $\Delta=0.9$eV relevant for MoTe$_2$. ${\bf b}$: Heterobilayer with velocity $v_t=2.2$eV\AA ~and $v_b=2.6$eV\AA, energy gap $\Delta_t=1.21$eV and $\Delta_b=1.16$eV representative values for WSe$_2$/MoSe$_2$. In both cases, we find spontaneous formation of excitons around $V^*_b\approx 70/75$meV. 
    The calculations were done with gate distance $\xi=12$nm and spacing between the two layers $d=1$nm.
    }
    \label{fig:TMDexcitons}
\end{figure}

We overlay this model with a gated Coulomb interaction
\begin{equation}
\mathcal{H}_{\rm int} = \dfrac{1}{2A} \sum^\Lambda_{{\bq}} \sum_{\nu,\nu^\prime=t,b}V^{\nu\nu^\prime}_{\bq} \hat{n}_{{\bq},\nu} \hat{n}_{-{\bq},\nu^\prime}
\end{equation}
where $\hat{n}_{{\bq},\nu}$ is the layer resolved density operator $\hat{n}_{{\bq},\nu}=\sum^\Lambda_{\bk}\sum_{a=K,K^\prime} \psi^\dagger_{{\bk},a\nu}\psi_{{\bk}+{\bq},a\nu}$ and
\begin{equation}
 V^{tt}_{\bq}=V^{bb}_{\bq}=\frac{ e^2}{\epsilon \epsilon_0}\frac{\tanh\frac{q\xi}{2}}{2q},\quad V^{tb}_{\bq}=V^{bt}_{\bq}=e^{-dq}V^{tt}_{\bq}.
\end{equation}
Here $d$ is the inter-layer distance, $\Lambda$ is the UV cutoff, and $\xi$ is the screening length of the bilayer which is defined as the distance between the bilayer and the metallic gates. 
As representative values~\cite{Ma2021,Shi_2022}, we set $\xi=12$nm, $\epsilon\approx6$ and $d\approx1$nm. 
We focus on the regime of low-density of electron-hole pairs \rev{$n_{\rm e-h}\le 0.075$} to reduce screening effects leading to the exciton Mott transition~\cite{mott1949basis, Mott1973, Brinkman_1973, Smith1986, Guerci2019}.
We now proceed with a Hartree-Fock calculation. 
We find that while the Hartree term vanishes as a consequence of charge neutrality, the Fock term gives rise to a self energy correction to the single-particle Hamiltonian. 
The self-consistent equations for the Fock self-energy are solved employing an iterative scheme, decribed in App.~D.

The Coulomb interaction induces intravalley excitons as well as charge transfer  between the two layers \rev{$\phi_z =\langle{\tau^z\sigma^0}\rangle=2n_{\rm e-h}$}. The operators $ \tau^{\pm}\sigma^a$ probe spontaneous interlayer coherence where $\tau^\pm = (\tau^x\pm i\tau^y)/2$ are raising/lowering operators in the layer degree of freedom and $\sigma^a$ are Pauli matrices in the band space.
A finite expectation value $\langle \mathcal \tau^{\pm}\sigma^a\rangle\neq 0$ corresponds to an interlayer intravalley exciton condensate that breaks the $U(1)_{e/h}$ symmetry. 
The first panel of Fig.~\ref{fig:TMDexcitons} shows the evolution of $\phi=\sqrt{\sum_{a=x,y,z}\langle \mathcal \tau^{x}\sigma^{a} \rangle^2}$ in the gauge where interlayer symmetry is broken along $\tau^x$ as a function of $V_b$ for the homobilayer (Fig.~\ref{fig:TMDexcitons}{\bf a}) and heterobilayer (Fig.~\ref{fig:TMDexcitons}{\bf b}) case. 
We notice that for the continuum theory the energy $V_b$ is given by $V_b=(\Delta_t+\Delta_b-2E_z)/2$ with $E_z$ electric displacement field applied to the bilayer. 
Above a critical value of the electrostatic potential $V_b^*\approx 70/75$meV the system turns into a semiconductor where the energy gap $E_{\text{gap}}$ grows linearly with the applied bias.
Finally, the phase stiffness is calculated using Eq.~\eqref{eq:Ds-Multi-Band} where the expectation value is taken with the Hartree-Fock wavefunctions. 
The results are shown in the right panel of Fig.~\ref{fig:TMDexcitons}{\bf a}-{\bf b} blue data for parabolic bands and green ones for a gapped Dirac cone. 
We emphasize that the non-trivial structure of the wavefunctions appears in the geometric contribution of the superfluid stiffness, while it does not change the exciton binding energy (energy gap) and the size of the order parameter. 

Remarkably, the geometric component increases $D_s$ and, correspondingly, the BKT transition temperature $(T_{\rm BKT})$ related to the superfluid stiffness by the Nelson-Kosterlitz relation \rev{ $k_BT_{\rm BKT}= \pi/2 (\lim_{T \rightarrow T_{\rm BKT}^-} D_s(T))$}~\cite{Nelson1977}. 
Approximating \rev{$D_s(T)$} with its $T=0$ value (from last panel in Fig.~\ref{fig:TMDexcitons}{\bf b}), we infer that geometric contribution increases $T_{\rm BKT}$ roughly from 12 to 36 Kelvin at $V_b \approx 40 $meV. This threefold increase should be observable in experiments.

\paragraph*{Conclusion.---}
Phase stifness is significantly modified by the constituent electron and hole wavefunctions of the exciton. 
We investigated two limiting scenarios, one with local Hubbard interaction and the other with gated Coulomb interaction projected onto the low-energy bands, and remarkably, we find geometric contributions that play a vital role in either case.
Our findings complement existing research on the effects of quantum geometry in exciton spectrum \cite{Zhou2015, Srivastava2015, Yao2008}, exciton wavefunction \cite{Sethi2023} and its possible realizations in moire heterostructures \cite{kwan2021,Hu2022, Hu2023}.
Importantly, we find a novel gauge-invariant quantum geometric quantity $\mathcal{G}({\bf k})$ whose properties require further investigation.

Bilayer TMD devices offer an ideal platform to validate our predictions.
\rev{ Quite recently, there have been two reports on exciton superfluidity in WSe$_2$/MoSe$_2$ bilayer \cite{Nguyen2023,Qi2023}. 
The BKT temperature of $20$ K appears to be consistent with the quantum geometric contribution, although a more sophisticated calculation is needed for a more quantitative prediction.}
Critical counterflow current from non-linear transport, similar to TBG superconductor \cite{Tian2023}, may also provide clues for the said geometric contribution.

\paragraph*{Acknowledgment.---}
Research on geometric properties of exciton condensates is supported as part of Programmable Quantum Materials, an Energy Frontier Research Center funded by the U.S. Department of Energy (DOE), Office of Science, Basic Energy Sciences (BES), under award DE-SC0019443. 
We benefited from the 2023 Quantum Geometry Working Group meeting that took place at the Flatiron institute.
We thank Päivi Törmä and Jonah Herzog-Arbeitman for their comments on orbital embeddings and minimal quantum metric.
We acknowledge discussions with Kin Fai Mak, Pavel Volkov, Yongxin Zeng and Ilia Komissarov.
D.G. and R.Q. acknowledge support from the Flatiron Institute, a division of the Simons Foundation.


%


\clearpage
\newpage

\begin{appendix}

\onecolumngrid
\newpage
\makeatletter 

\begin{center}
\textbf{\large Supplementary material for ``\@title ''}

\vspace{10pt}

Nishchhal Verma,$^{1,*}$ Daniele Guerci,$^{2,*}$ and Raquel Queiroz$^{1,2}$ 

\textit{${}^1$Department of Physics, Columbia University, New York, NY 10027, USA}

\textit{${}^2$Center for Computational Quantum Physics, Flatiron Institute, New York, New York 10010, USA} 

\end{center}
\vspace{20pt}

\setcounter{figure}{0}
\setcounter{section}{0}
\setcounter{equation}{0}

\renewcommand{\thefigure}{S\@arabic\c@figure}
\renewcommand{\thesection}{S-\@Roman\c@section}
\makeatother

\appendix

\section{Mean-field Phase Stiffness}\label{app:Ds-mf}
\subsection{Kubo formula}\label{app:Ds-multi-band}
Exciton condensates exhibit the remarkable phenomenon of dissipationless counterflow transport when equal and opposite fields are applied to the two layers.
The asymmetric field drives excitons in a coherent fashion.
The response kernel to layer anti-symmetric vector potential is defined as \cite{Eisenstein2014}
\begin{equation}
    j_{ {\rm CF} } = -\dfrac{4e^2}{\hbar^2} D_s A
\end{equation}    
where $D_s$ is the phase stiffness and ${\bf A}=A \hat{x}$ is the external vector potential that is applied along $+\hat{x}$ in top layer and $-\hat{x}$ in bottom layer.
As a matter of fact, this can be an in-plane magnetic field as ${\bf A}=zA\hat{x}$ gives $B_y = \partial_z A_x$ where $z$ is the direction normal to the interface \cite{Sun2021}.
The stiffness is the weight of the delta function in counterflow conductivity ${\rm Re}[\sigma_{ {\rm CF} }(\omega)] = D_s \delta(\omega) + \cdots$.
By utilizing the Kramers-Kronig relation, $D_s$ can be extracted from the $1/\omega$ tail of ${\rm Im}[\sigma_{CF}(\omega)]$, which follows ${\rm Im}[\sigma_{CF}(\omega)] \propto D_s/\omega$.
Stiffness controls the cost of phase fluctuations of the condensate and is formally defined by the Kubo formula \cite{scalapino1993}
\begin{equation}
    D_s = -\dfrac{1}{4} \Big( \langle j_D \rangle - \chi_{ j_P j_P}({\bf q}_\perp = 0, \omega=0) \Big) .\label{app:eq:ds:kubo}
\end{equation}
The diamagnetic $(j_D)$ and paramagnetic $(j_P)$ currents are given by a Taylor expansion of the Hamiltonian
\begin{eqnarray}
    \mathcal{H}_{\rm ex}(A) &=& \sum\limits_{ {\bf k} } \Psi^\dag_{ {\bf k} } \left[ \begin{pmatrix}
        H_t({\bf k} - eA/\hbar) & 0 \\ 0 & H_b({\bf k} + eA/\hbar)
    \end{pmatrix} + V_b \tau^z \right] \Psi^\dag_{ {\bf k} } + \mathcal{H}_{\rm int} \\
    &=& \mathcal{H}_{\rm ex}(0) - \dfrac{eA}{\hbar} \sum\limits_{{\bf k}} \Psi^\dag_{ {\bf k} } \begin{pmatrix} \partial_k H_t({\bf k}) & 0 \\ 0 & -\partial_k H_b({\bf k}) \end{pmatrix} \Psi^\phdag_{ {\bf k} } + \dfrac{e^2A^2}{2\hbar^2} \sum\limits_{{\bf k}} \Psi^\dag_{ {\bf k} } \begin{pmatrix} \partial_k^2 H_t({\bf k}) & 0 \\ 0 & \partial_k^2 H_b({\bf k}) \end{pmatrix} \Psi^\phdag_{ {\bf k} }
\end{eqnarray}
which gives
\begin{eqnarray}
    j_P = \dfrac{e}{\hbar} \sum\limits_{{\bf k}} \Psi^\dag_{ {\bf k} } \begin{pmatrix} \partial_k H_t({\bf k}) & 0 \\ 0 & -\partial_k H_b({\bf k}) \end{pmatrix} \Psi^\phdag_{ {\bf k} } , \quad j_D = -\dfrac{e^2}{\hbar^2} \sum\limits_{{\bf k}} \Psi^\dag_{ {\bf k} } \begin{pmatrix} \partial_k^2 H_t({\bf k}) & 0 \\ 0 & \partial_k^2 H_b({\bf k}) \end{pmatrix} \psi^\phdag_{ {\bf k} } \label{app:eq:curr-ops}.
\end{eqnarray}
Next, in order to calculate the expectation values in Eq.~\eqref{app:eq:ds:kubo}, we assume a mean-field Hamiltonian with eigenspectrum $E_{m,{\bf k}}$ and eigenstates $| u_{m,{\bf k}}\rangle \}$ to find
\begin{equation}
    \langle j_D \rangle = \dfrac{1}{A}\sum\limits_{ {\bf k}, m} f[E_{m,{\bf k}}] \left\langle u_{m, {\bf k}} \left| \partial^2_k H_0({\bf k}) \right| u_{m, {\bf k}} \right\rangle, \quad    \chi_{ j_P j_P} = \dfrac{1}{A}\sum\limits_{{\bf k}, m\neq n} \dfrac{f[E_{m,{\bf k}}]-f[E_{n,{\bf k}}]}{E_{n,{\bf k}}-E_{m,{\bf k}}}  \left|\left\langle u_{m, {\bf k}} \left| \tau^z \partial_k H_0({\bf k}) \right| u_{n, {\bf k}} \right\rangle\right|^2
\end{equation}
which gives Eq.~(1) of the main text.
The expression not only captures the conventional contribution coming from energies but also includes all geometric contributions that are hidden inside the matrix elements.
These geometric factors arise from the underlying electron and hole wavefunctions that are not directly visible in this equation.
We will now describe a projected mean-field model which systematically uncovers these geometric contributions. 

\subsection{Projected Phase Stiffness}\label{app:Ds-proj-band}
If the electron and hole bands are sufficiently isolated from other bands, we can focus on the low-energy Hamiltonian and couple to the external vector potential
\begin{equation}
\mathcal{H}_{\rm ex}(A) = \sum\limits_{ {\bf k}} \psi^\dag_{\bf k} \begin{pmatrix}
         \epsilon({\bf k}-eA/\hbar) - \Sigma({\bf k}) & \varphi({\bf k},A) \\  \phi \;\varphi({\bf k},A)^* & -\Big( \epsilon({\bf k}+eA/\hbar) - \Sigma({\bf k}) \Big)
    \end{pmatrix} \psi^\phdag_{ {\bf k}} = \sum\limits_{ {\bf k}} \psi^\dag_{\bf k} H_{\rm ex}(A) \psi^\phdag_{ {\bf k}}.
\end{equation}

The vector potential dependence on the off-diagonal component is key to the geometric contribution
\begin{eqnarray}
    \varphi({\bf k}, A) &=& \sum_{ \alpha\beta} [U_t({\bf k}-eA/\hbar)]^*_{e, \alpha} \phi_{\alpha\beta}({\bf k}) [U_b({\bf k}+eA/\hbar)]_{\beta,h} = \varphi({\bf k}) - \dfrac{e A}{\hbar} \mathcal{P}({\bf k}) - \dfrac{e^2 A^2}{2\hbar^2} \mathcal{D}({\bf k})
\end{eqnarray}
where $\mathcal{P}({\bf k})$ and $\mathcal{D}({\bf k})$ are defined as
\begin{eqnarray}
    \mathcal{P}({\bf k}) = \sum_{ \alpha\beta} \partial_k [U_t({\bf k})]^*_{e, \alpha} \phi_{\alpha\beta}({\bf k}) [U_b({\bf k})]_{\beta, h} - [U_t({\bf k})]^*_{e, \alpha} \phi_{\alpha\beta}({\bf k}) \partial_k [U_b({\bf k})]_{\beta, h} \hspace{5.45cm}  \\
    \mathcal{D}({\bf k}) = \sum_{ \alpha\beta} 2 \partial_k [U_t({\bf k})]^*_{e, \alpha} \phi_{\alpha\beta}({\bf k}) \partial_k [U_b({\bf k})]_{\beta, h} - \partial_k^2[U_t({\bf k})]^*_{e, \alpha} \phi_{\alpha\beta}({\bf k}) [U_b({\bf k})]_{\beta, h} - [U_t({\bf k})]^*_{e, \alpha} \phi_{\alpha\beta}({\bf k}) \partial_k^2 [U_b({\bf k})]_{\beta, h} .
\end{eqnarray}

Moving onto the stiffness calculation, we will use the thermodynamic definition instead of using the Kubo formula.
Focusing on $T=0$ and assuming that the mean-field band is completely gapped (even for $H_{\rm ex}({\bf k},A)$), the stifness is given by
\begin{equation}
    D_s = \dfrac{\hbar^2}{4 e^2 } \left( \dfrac{\partial^2} {\partial A^2} \mathcal{F}(A)\right) = \dfrac{\hbar^2}{4 e^2 } \sum\limits_{ {\bf k} } \dfrac{\partial^2} {\partial A^2} \mathcal{E}_-({\bf k}, (A) )
\end{equation}
where $\mathcal{E}_-({\bf k}, A )$ is the lower band corresponding to $H_{\rm ex}({\bf k},A)$, given by 
\begin{equation}
\mathcal{E}_\pm({\bf k}, A ) = \xi_-({\bf k},A) \pm \sqrt{ \xi_+({\bf k},A)^2 + | \varphi({\bf k},A) |^2}    
\end{equation}
where
\begin{equation}
    \xi_\pm({\bf k},A) = \dfrac{1}{2} \Big[ ( \epsilon({\bf k}-eA/\hbar) - \Sigma({\bf k}) ) \pm ( \epsilon({\bf k} + eA/\hbar) - \Sigma({\bf k}) ) \Big]
\end{equation}
that has convenient properties like
\begin{eqnarray}
    \xi_-({\bf k},0) = 0, &\quad & \xi_+({\bf k},0) = \epsilon({\bf k}) - \Sigma({\bf k}) \equiv \xi({\bf k}) \\
    \partial_A\xi_-({\bf k},A)|_{A=0} = -\dfrac{e}{\hbar} \partial_k \epsilon({\bf k}), &\quad &\partial_A\xi_+({\bf k},A)|_{A=0} = 0 \\
    \partial_A^2 \xi_-({\bf k},A)|_{A=0} = 0, & \quad & \partial_A^2 \xi_+({\bf k},A)|_{A=0} = \dfrac{e^2}{\hbar^2} \partial_k^2\epsilon({\bf k}).
\end{eqnarray}
We are now ready to evaluate $D_s$:
\begin{eqnarray}
    D_s &=& \dfrac{\hbar^2}{4e^2} \dfrac{\partial^2}{\partial A^2} \left( \sum\limits_{\bf k} \mathcal{E}_-({{\bf k}},A) \right)_{A=0} = \dfrac{\hbar^2}{4e^2} \dfrac{\partial^2}{\partial A^2} \left( \sum\limits_{\bf k} ( \xi_-({\bf k},A) - \sqrt{ \xi_+({\bf k},A)^2 + | \varphi({\bf k},A) |^2  } \right)_{A=0} \\
    &=& \dfrac{\hbar^2}{4e^2} \sum\limits_{\bf k} \partial_A^2 \xi_-({\bf k},A)|_{A=0} - \dfrac{\partial^2}{\partial A^2} \left( \sqrt{ \xi_+({\bf k},A)^2 + | \varphi({\bf k},A) |^2  } \right)_{A=0} \\
    &=& - \dfrac{1}{4} \sum\limits_{\bf k} \partial_k^2 \epsilon({\bf k}) \dfrac{ \xi({\bf k})}{E({\bf k})} +  \dfrac{\hbar^2}{8e^2}\sum\limits_{\bf k} \dfrac{(-\partial_A^2 | \varphi({\bf k},A) |^2) }{ E({\bf k}) } + \dfrac{\hbar^2}{4e^2} \sum\limits_{\bf k}\dfrac{ (\partial_A | \varphi({\bf k},A) |^2/2)^2 }{ E({\bf k})^3 } .
\end{eqnarray}
Using the fact that $\sum_{\bf k}\partial_k^2\epsilon({\bf k}) = 0$, we add and subtract this term to write 
\begin{equation}
    D_s = \dfrac{1}{4} \sum\limits_{\bf k}\partial_k^2 \epsilon({\bf k}) \left[ 1 - \dfrac{\xi({\bf k})}{ E({\bf k}) } \right] +  \dfrac{\hbar^2}{8e^2}\sum\limits_{\bf k} \dfrac{(-\partial_A^2 | \varphi({\bf k},A) |^2) }{ E({\bf k}) } + \dfrac{\hbar^2}{4e^2} \sum\limits_{\bf k}\dfrac{ (\partial_A | \varphi({\bf k},A) |^2/2)^2 }{ E({\bf k})^3 }
\end{equation}
where we can further use
\begin{equation}
    | \varphi({\bf k},A) |^2 = |\varphi({\bf k})|^2 -2 \dfrac{e A}{\hbar} {\rm Re}[ [\mathcal{P}({\bf k}) \varphi({\bf k})^* ] - \dfrac{e^2 A^2}{\hbar^2} \Big( {\rm Re}[ [\mathcal{D}({\bf k}) \varphi({\bf k})^* ] - | \mathcal{P}({\bf k}) |^2  \Big) \label{eq:varphiAexpand}
\end{equation}
to arrive at the result quoted in the main text
\begin{equation}
    D_s = \dfrac{1}{2} \sum\limits_{\bf k} \partial_k^2 \epsilon({\bf k}) v_{\bf k}^2 + \dfrac{1}{4}\sum\limits_{\bf k} \dfrac{ {\rm Re }[\mathcal{D}({\bf k}) \varphi({\bf k})^*] - |\mathcal{P}({\bf k})|^2 }{ E({\bf k}) } + \dfrac{1}{4}\sum\limits_{\bf k} \dfrac{ {\rm Re}[\mathcal{P}({\bf k}) \varphi({\bf k})^* ]^2 }{ E({\bf k})^3 }.\label{eq:Ds-app-full}
\end{equation}

\subsection{Reduction to Quantum Metric}\label{app:Ds-qgt}
We can validate our calculations by making two assumptions: first,$\hat{\phi}({\bf k}) = \phi \mathbb{1},\; \phi \in \mathbb{R}$ and second, the two layers are identical, giving $[U_t({\bf k})]_{e,\alpha} = [U_b({\bf k})]_{h,\alpha} \;\forall\; \alpha, {\bf k}$. 
We can express the matrix elements in a more concise form (dropping $e$, $h$ labels):
\begin{eqnarray}
\varphi({\bf k}) = \phi, \quad \mathcal{P}({\bf k}) = 2\phi \sum_{ \alpha} \partial_k U^*_{\alpha}({\bf k}) U_{\alpha}({\bf k}) \equiv 2\phi \langle \partial_k u_{\bf k} | u_{\bf k}\rangle, \quad    \mathcal{D}({\bf k}) = 4 \phi \sum_{ \alpha} \partial_k U^*_{ \alpha}({\bf k}) \partial_k U_{\alpha}({\bf k}) \equiv 4\phi \langle \partial_k u_{\bf k} | \partial_k u_{\bf k}\rangle 
\end{eqnarray}
and the resulting geometric contribution to stiffness greatly simplifies
\begin{eqnarray}
    D_s^g &=& \dfrac{1}{4A} \sum\limits_{\bf k} \dfrac{1}{ E({\bf k})} \left( {\rm Re}[\mathcal{D}({\bf k}) \varphi({\bf k})^*] -  \left| {\rm Im}[ \mathcal{P}({\bf k}) ]  \right|^2 \right) = \dfrac{1}{A}\sum\limits_{\bf k} \dfrac{\phi^2}{ E({\bf k})} \left( {\rm Re}\left[\langle \partial_k u_{\bf k} | \partial_k u_{\bf k}\rangle\right] -  \left| \langle \partial_k u_{\bf k} | u_{\bf k}\rangle \right|^2 \right) \\
    &=& \dfrac{1}{A}\sum\limits_{\bf k} \dfrac{\phi^2}{ E({\bf k})} {\rm Re}[\mathcal{Q}({\bf k})].
\end{eqnarray}
Here ${\rm Re}[\mathcal{Q}({\bf k})] = g({\bf k})$ is the quantum metric of the electron (or hole) band.

\subsection{Gauge Invariance}\label{app:gaugeInvariance}
There is a gauge redundancy in non-interacting electron/hole wavefunctions since Bloch wavefunctions are only defined upto a phase.
Physical quantities must be invariant under such gauge transformations.
Specifically, introducing phases in the wavefunctions should not modify any observable
\begin{equation}
    U_t({\bf k})_{\alpha, e} \rightarrow e^{i \phi_t({\bf k}) } U_t({\bf k})_{\alpha, e}, \quad U_b({\bf k})_{\beta, h} \rightarrow e^{i \phi_b({\bf k}) } U_b({\bf k})_{\beta, h}
\end{equation}
where $\phi_\nu({\bf k})$ are real but arbitrary functions in the BZ.
It turns out that $\varphi({\bf k}), \mathcal{P}({\bf k})$ and $\mathcal{D}({\bf k})$ are sensitive to the gauge choice
\begin{eqnarray}
    \varphi({\bf k}) &\rightarrow & e^{ -i (\phi_t({\bf k}) - \phi_b({\bf k}) ) } \varphi({\bf k}) \\
    \mathcal{P}({\bf k}) &\rightarrow & e^{ -i (\phi_t({\bf k}) - \phi_b({\bf k}) ) } \Big[ \mathcal{P}({\bf k}) - i \varphi({\bf k})[ \partial_k \phi_t({\bf k}) + \partial_k \phi_b({\bf k}) ] \Big] \\
    \mathcal{D}({\bf k}) &\rightarrow & e^{ -i (\phi_t({\bf k}) - \phi_b({\bf k}) ) } \Big[ \mathcal{D}({\bf k}) - \varphi({\bf k}) [ \partial_k \phi_t({\bf k}) + \partial_k \phi_b({\bf k}) ]^2 + 2i [ \partial_k \phi_t({\bf k}) + \partial_k \phi_b({\bf k}) ] \mathcal{P}({\bf k}) \Big].
\end{eqnarray}
However, the combinations that enter geometric contributions to phase stiffness are gauge invariant.
Eq.~\eqref{eq:varphiAexpand} offers a simple proof.
The left-hand side is gauge invariant and hence the right-hand side has to be gauge invariant for arbitrary expansions in $A$.

It can also be checked by an explicit calculation.
The first term, ${\rm Re}[\mathcal{P}({\bf k})\varphi({\bf k})^*]$, gets an additional term which is purely imaginary
\begin{eqnarray}
   \mathcal{P}({\bf k})\varphi({\bf k})^* \rightarrow \mathcal{P}({\bf k})\varphi({\bf k})^* - i |\varphi({\bf k})|^2 [ \partial_k \phi_t({\bf k}) - \partial_k \phi_b({\bf k}) ] 
\end{eqnarray}
and hence ${\rm Re}[\mathcal{P}({\bf k})\varphi({\bf k})^*]$ is gauge invariant.
The second term follows along similar lines
\begin{eqnarray}
    \mathcal{D}({\bf k}) \varphi({\bf k})^* &\rightarrow &  \mathcal{D}({\bf k}) \varphi({\bf k})^* - |\varphi({\bf k})|^2 [ \partial_k \phi_t({\bf k}) +  \partial_k \phi_b({\bf k})]^2 + 2i \mathcal{P}({\bf k}) \varphi({\bf k})^* [ \partial_k \phi_t({\bf k}) +  \partial_k \phi_b({\bf k})] \\
    |\mathcal{P}({\bf k})|^2 &\rightarrow &  |\mathcal{P}({\bf k})|^2 + |\varphi({\bf k})|^2 [ \partial_k \phi_t({\bf k}) +  \partial_k \phi_b({\bf k})]^2 - 2 {\rm Im}[\mathcal{P}({\bf k}) \varphi({\bf k})^*] [ \partial_k \phi_t({\bf k}) +  \partial_k \phi_b({\bf k})] 
\end{eqnarray}
where the last two terms exactly cancel and ${\rm Re}[\mathcal{D}({\bf k}) \varphi({\bf k})^* ] - |\mathcal{P}({\bf k})|^2$ is gauge invariant.

\subsection{Orbital Embedding }\label{app:minimalQM}
In addition to the gauge freedom mentioned above, there is an another transformation that changes the wavefunctions, but does not affect the energies
\begin{equation}
    U_t({\bf k})_{\alpha, e} \rightarrow e^{i \phi_{t, \alpha} ({\bf k}) } U_t({\bf k})_{\alpha, e}, \quad U_b({\bf k})_{\beta, h} \rightarrow e^{i \phi_{b,\beta}({\bf k}) } U_b({\bf k})_{\beta, h}
\end{equation}
where the phases depend on orbital index $\alpha$ (and $\beta$). 
The transformation corresponds to physically moving the orbitals within the unit cell while preserving all hopping amplitudes.
It changes Berry curvature \cite{Simon2020}, quantum metric \cite{Huhtinen2022} and the geometric contribution to stiffness.
It has been argued that orbital embedding should not change the phase stiffness \cite{Peotta2022} and the correct prescription is discussed in detail in ref.~\cite{Huhtinen2022}.
The gauge physically corresponds to orbital locations that keep order parameter solutions real in the presence of vector potential.

\subsection{Spatially Anisotropic Systems}\label{app:anisotropicSystem}
Underneath all our discussion so far, we have assumed isotropic systems where the vector potential is applied along $\hat{x}$ with ${\bf A} = A \hat{x}$ and the longitudinal response is calculated along $\hat{x}$ while suppressing all spatial indices. 
In this section, we will outline the extension to spatially anisotropic systems.
Introducing indices $i,j = \{ x, y\}$, we find that the phase stiffness in Eq.~(A6) becomes a tensor
\begin{eqnarray}
[D_s]_{ij} &=& -\dfrac{1}{4A} \Big( \langle [j_D]_{ij} \rangle - \chi_{ [j_P]_i [j_P]_j}({\bf q}_\perp=0, \omega=0) \Big) \\
&=& \dfrac{\hbar^2}{4e^2A} \Big[ \sum\limits_{ {\bf k}, m} f[E_{m,{\bf k}}] \left\langle u_{m, {\bf k}} \left|  [\hat{j}_D({\bf k})]_{ij} \right| u_{m, {\bf k}} \right\rangle \nonumber  \\
&\phantom{.} & \quad \quad \quad \quad - \sum\limits_{{\bf k}, m\neq n} \dfrac{f[E_{m,{\bf k}}]-f[E_{n,{\bf k}}]}{E_{n,{\bf k}}-E_{m,{\bf k}}} \left\langle u_{m, {\bf k}} \left| [\hat{j}_P({\bf k})]_i \right| u_{n, {\bf k}} \right\rangle \left\langle u_{n, {\bf k}} \left| [\hat{j}_P({\bf k})]_j \right| u_{m, {\bf k}} \right\rangle \Big]
\end{eqnarray}
where the current operators are
\begin{eqnarray}
    [\hat{j}_P({\bf k})]_i = \dfrac{e}{\hbar}\begin{pmatrix} \partial_{k_i}H_t({\bf k}) & 0 \\ 0 & -\partial_{k_i}H_b({\bf k}) \end{pmatrix}, \quad [\hat{j}_D({\bf k})]_{ij} = -\dfrac{e^2}{\hbar^2} \begin{pmatrix} \partial_{k_i}\partial_{k_j}H_t({\bf k}) & 0 \\ 0 & \partial_{k_i}\partial_{k_j}H_b({\bf k}) \end{pmatrix} \label{app:eq:curr-ops}.
\end{eqnarray}
The projected stiffness in Eq.~\eqref{eq:Ds-app-full} follows along similar lines
\begin{eqnarray}
    [D_s]_{ij} = \dfrac{1}{4A}\sum\limits_{\bf k} \left[ \partial_{k_i} \partial_{k_j} \epsilon({\bf k}) \left( 1- \dfrac{\xi({\bf k})}{ E({\bf k}) }\right) +  \sum\limits_{\bf k} \dfrac{[\mathcal{G}({\bf k})]_{ij}}{ E({\bf k})} + \sum\limits_{\bf k} \dfrac{{\rm Re}[ [\mathcal{P}({\bf k})]_i \varphi({\bf k})^*] {\rm Re}[ [\mathcal{P}({\bf k})]_j \varphi({\bf k})^*]}{ E({\bf k})^2} \right]
\end{eqnarray}
where $[\mathcal{G}({\bf k})]_{ij} = {\rm Re}[\mathcal{D}({\bf k})]_{ij}\varphi({\bf k})^*] -  [\mathcal{P}({\bf k})]_i [\mathcal{P}({\bf k})]_j^* $ and the functions $[\mathcal{P}({\bf k})]_{i}$, $[\mathcal{D}({\bf k})]_{ij}$ are
\begin{eqnarray}
    [\mathcal{P}({\bf k})]_{i} &=& \sum_{ \alpha\beta} \Big[ \partial_{k_i} [U_t({\bf k})]^*_{e, \alpha} \phi_{\alpha\beta}({\bf k}) [U_b({\bf k})]_{\beta, h} - [U_t({\bf k})]^*_{e, \alpha} \phi_{\alpha\beta}({\bf k}) \partial_{k_i} [U_b({\bf k})]_{\beta, h} \Big]
\end{eqnarray}
\begin{eqnarray}
    [\mathcal{D}({\bf k})]_{ij} &=& \sum_{ \alpha\beta} \Big[ \partial_{k_i} [U_t({\bf k})]^*_{e, \alpha} \phi_{\alpha\beta}({\bf k}) \partial_{k_j} [U_b({\bf k})]_{\beta, h} + \partial_{k_j} [U_t({\bf k})]^*_{e, \alpha} \phi_{\alpha\beta}({\bf k}) \partial_{k_i} [U_b({\bf k})]_{\beta, h} \nonumber \\
    &\phantom{\quad }& - \partial_{k_{i}}\partial_{k_{j}} [U_t({\bf k})]^*_{e, \alpha} \phi_{\alpha\beta}({\bf k}) [U_b({\bf k})]_{\beta, h} - [U_t({\bf k})]^*_{e, \alpha} \phi_{\alpha\beta}({\bf k}) \partial_{k_{i}}\partial_{k_{j}} [U_b({\bf k})]_{\beta, h} \Big].
\end{eqnarray}
Lastly, the BKT temperature is determined by the determinant of the stiffness tensor $k_B T_c = (\pi/2) \sqrt{ {\rm det} [D_s(T_c^-)]_{ij} }$.

\section{Mean-Field theory for the Hubbard interaction}\label{app:sec:MeanField}
We consider an interlayer on-site orbital-diagonal Hubbard interaction  $\mathcal{H}_{\rm int} = V \sum_{{\bf i},\alpha} \hat{n}_{{\bf i}, \alpha, t} \hat{n}_{{\bf i}, \alpha, b}$.
The labels ${\bf i}, \alpha$ pertaining to unit cell index and orbitals, are not crucial for our discussion as long as the interaction is interlayer in character.
We decouple the four-fermion interaction with a spatially uniform ansatz
\begin{eqnarray}
V_{\alpha\beta} \;c^\dag_{ {\bf i}, \alpha, t} c^\phdag_{ {\bf i}, \alpha, t} \; c^\dag_{ {\bf i}, \beta, b} c^\phdag_{ {\bf i}, \beta, b} &\approx & V_{\alpha\beta} \;n_{ \alpha, t}  c^\dag_{ {\bf i}, \beta, b} c^\phdag_{ {\bf i}, \beta, b} + V_{\alpha\beta} c^\dag_{ {\bf i}, \alpha, t} c^\phdag_{ {\bf i}, \alpha, t}  n_{ \beta, b} + \phi_{\alpha\beta}  c^\dag_{ {\bf i}, \beta, b} c^\phdag_{ {\bf i}, \alpha, t} + \phi_{\alpha\beta}^\dag  c^\dag_{ {\bf i}, \alpha, t} c^\phdag_{ {\bf i}, \beta, b}
\end{eqnarray}
where $\{ n_{ \alpha, \nu} \}$ and $\{\phi_{\alpha\beta}\}$ are given by
\begin{equation}
    n_{ \alpha, \nu} = \dfrac{1}{A} \sum\limits_{ {\bf i} } \left\langle c^\dag_{ {\bf i}, \alpha, \nu} c^\phdag_{ {\bf i}, \alpha, \nu}\right\rangle = \dfrac{1}{N} \sum\limits_{ {\bf k} } \left\langle c^\dag_{ {\bf k}, \alpha, \nu} c^\phdag_{ {\bf k}, \alpha, \nu}\right\rangle, \quad \phi_{\alpha\beta} = - \dfrac{V_{\alpha\beta}}{A} \sum\limits_{{\bf i}} \left\langle c^{\dag}_{ {\bf i}, \alpha, t} c^{\phantom\dag}_{ {\bf i}, \beta, b} \right\rangle = - \dfrac{V_{\alpha\beta}}{N} \sum\limits_{{\bf k}} \left\langle c^{\dag}_{ {\bf k}, \alpha, t} c^{\phantom\dag}_{ {\bf k}, \beta, b} \right\rangle. \label{app:eq:selfCons-def}
\end{equation}
Here $A = N \Omega$ is the area of the system, $N$ is the number of unit cells and $\Omega$ is the area of unit cell.
This pre-factor serves as a normalization factor. 
When we substitute this result back into the full Hamiltonian, we obtain the Hartree shifts
\begin{eqnarray}
    [\widetilde{H}(\zeta_t, {\bf k})]_{\alpha \alpha^\prime} = [H(\zeta_t, {\bf k})]_{\alpha \alpha^\prime} +  \delta_{\alpha,\alpha^\prime} V_{\alpha\beta} n_{\beta, b}, \quad [\widetilde{H}(\zeta_b, {\bf k})]_{\beta \beta^\prime} = [H(\zeta_b, {\bf k})]_{\beta \beta^\prime} +  \delta_{\beta,\beta^\prime} V_{\alpha\beta} n_{\alpha, t}
\end{eqnarray}
in addition to the interlayer exciton term $\propto \phi_{\alpha\beta}  c^\dag_{ {\bf i}, \beta, b} c^\phdag_{ {\bf i}, \alpha, t}$.
Combining the two, we get the mean-field Hamiltonian $\mathcal{H}_{\rm MF} = \sum_{{\bf k}} \Psi^\dag_{{\bf k}} H_{\rm MF}({\bf k}) \Psi^{\phantom\dag}_{{\bf k}}$ where
\begin{equation}
    H_{\rm MF}({\bf k}) = [\widetilde{H}(\zeta_t, {\bf k}) \oplus (-\widetilde{H}(\zeta_b, {\bf k}))] + \tau^z V_b  + \tau^x \phi .\label{eq:meanFieldFullHam}
\end{equation}
The last term breaks the $U(1)_t \times U(1)_b$ symmetry as discussed before.

To establish the notation, let us review the simplest model consisting of two identical bands with mass $m$: one particle-like and one hole-like, that belong to two different layers. 
We consider the system at charge neutrality, which means that the sum of the particle density $n_t$ and the hole density $n_b$ equals 1, and introduce an inter-layer bias $V_b$.
The interaction term is $U \hat{n}_t (\hat{n}_b-1)$, specifically chosen to vanish when the lower band is fully occupied ($n_b=1$).
We then decouple the interaction into two components: a real exciton field denoted by $\phi$, which can be expressed as $\phi = -U (\langle c^\phdag_{b} c^\phdag_{t} \rangle + \langle c^\dag_{t} c^\dag_{b} \rangle)/2$, and an inter-layer polarization field denoted by $\phi_z$ which is $\phi_z = U \langle c^\dag_{t} c^\phdag_{t} \rangle$.
Combining these, we obtain the mean-field Hamiltonian
\begin{equation}
    \mathcal{H}_{MF} = \sum\limits_{ {\bf k} } \psi^\dag_{{\bf k}} \Big[ \Big(\dfrac{{\bf k}^2}{2m} + V_b - \phi_z \Big) \tau_z + \phi \tau_x \Big] \psi^\phdag_{{\bf k}}.\label{eq:MFpb}
\end{equation}
where $\psi^\phdag_{\bf k} = ( c^\phdag_{e, {\bf k}}, c^\phdag_{h, {\bf k}} )^T$ (since electron/hole labels are identical to the layer labels).
Next, we define $\xi_{{\bf k}} = {\bf k}^2/2m + h - \phi_z$ and $E_{{\bf k}} = \sqrt{\xi_{{\bf k}}^2 + \phi^2}$ to write the self-consistent equations (at $T=0$) for $\phi$ and $\phi_z$ 
\begin{equation}
    \dfrac{1}{V} = \dfrac{1}{N}\sum\limits_{{\bf k}} \dfrac{1}{2E_{{\bf k}}}, \quad \phi_z = \dfrac{V}{2N} \sum\limits_{{\bf k}} \left(1 - \dfrac{\xi_{{\bf k}}}{E_{{\bf k}}}\right).\label{eq:gap-eq-pb}
\end{equation}
The solutions can be obtained numerically with a lattice regularization ${\bf k}^2/2m \rightarrow \epsilon_{{\bf k}} = 2t(2-\cos k_x-\cos k_y)$ where $t$ can be thought of as a nearest-neighbor hopping amplitude on a square lattice.

\begin{figure*}
    \centering
    \includegraphics[width=0.95\textwidth]{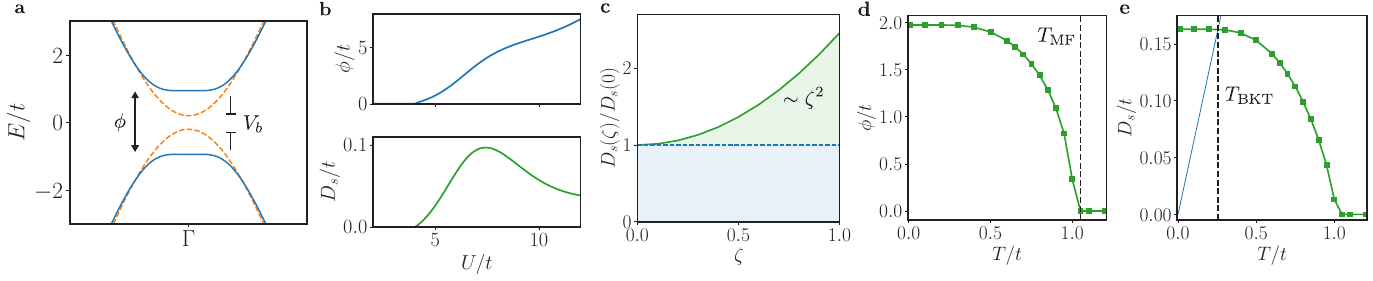}
    \caption{Mean-field theory for symmetric excitons in the minimal model. {\bf a} Both the particle and hole bands have the same dispersion and wavefunction. The non-interacting bands (dashed line) have a gap $V_b$ that increases further with the formation of excitons (solid line). {\bf b} Self-consistent solution of $\phi$ together with $D_s$. Both require a finite interaction strength $U/t > 0$ to activate. {\bf c} Geometric contribution to stiffness scales with $\zeta^2$ as expected from trace of the quantum metric. {\bf d} and {\bf e} show the temperature dependence of $\phi$ and $D_s$ respectively. 
    The mean-field temperature (at which $\phi$ vanishes) is clearly larger than the BKT temperature (intersection of $D_s(T)$ and $2T/\pi$). The condensate is limited by phase fluctuations.}
    \label{app:Fig:mean-field-parabolicBand}
\end{figure*}

The multi-band generalization follows along similar lines.
First we compute the eigenvalues and eigenvectors $\{ E_m({\bf k}), | u_{m, {\bf k}}\rangle\}$ of the mean-field Hamiltonian. 
The mean-field states follow the thermal occupation 
\begin{equation}
    \left\langle \gamma^\dag_{ {\bf k}, m } \gamma^\phdag_{ {\bf k}, m^\prime } \right\rangle= f[E_m({\bf k})]\delta_{m,m^\prime}
\end{equation}
where $f$ is the Fermi function and the operators $\{\gamma_{{\bf k}}\}$ are related to $\{c_{\bf k}\}$ by a unitary transformation.
We use the new operators $\{\gamma_{{\bf k}}\}$ to rewrite Eq.~\eqref{app:eq:selfCons-def} which turns out to have a concise matrix structure
\begin{equation}
    \phi_{\alpha\beta} = - \dfrac{V_{\alpha\beta}}{N} \sum\limits_{ {\bf k}} f[E_m({\bf k})] \left\langle u_{m, {\bf k}} \left| \dfrac{\partial \mathcal{H}_{MF}}{\partial \phi_{\alpha\beta}^*} \right| u_{m, {\bf k}} \right\rangle, \quad n_{\alpha,t} = \dfrac{V_{\alpha\beta}}{N} \sum\limits_{ {\bf k}} f[E_m({\bf k})] \left\langle u_{m, {\bf k}} \left| \dfrac{\partial \mathcal{H}_{MF}}{\partial n_{\beta,b}} \right| u_{m, {\bf k}} \right\rangle.
\end{equation}
The equation is dependent on $\{n_{\alpha,t}, \phi_{\alpha\beta}\}$ on both sides, thus requiring an iterative approach.
We start with a guess and use \texttt{FindRoot} function in Mathematica to find the solution.
The solutions are shown in Fig.~\ref{app:Fig:mean-field-parabolicBand}.

\subsection{Tunable Quantum Metric Model}
We reprint the Bloch hamiltonian for the tunable quantum metric model for convenience
\begin{equation}
    H(\zeta, {\bf k}) = 2t(2-p_{\bf k}) + t_F \cos(\zeta p_{\bf k}) \sigma^x + t_F \sin(\zeta p_{\bf k}) \sigma^y \label{eq:def:zeta-ham}.
\end{equation}
Here the Pauli matrices $\sigma^i$ act in orbital space, $p_{\bf k}$ is a periodic function given by $ p_{\bf k} = \cos k_x + \cos k_y$ and $\zeta$ is an overall scaling factor for the quantum geometric tensor $\mathcal{Q}( {\bf k}, \zeta )_{\mu\nu} = (\zeta^2/4) \sin k_\mu \sin k_\nu $.

We find a significant simplification with $\phi_{\alpha\beta}({\bf k}) = \phi \delta_{\alpha\beta}$.
This relatively simple form can be attributed to three key factors.
The on-site nature makes momentum independent $\hat{\phi}$ a natural ansatz. 
The orbital-diagonal aspect $\hat{n}_{\alpha,t} \hat{n}_{\alpha,b}$ removes the possibility of off-diagonal elements in $\hat{\phi}$.
Lastly, the two orbitals in the toy model are related by an in-plane inversion which ultimately reduces the exciton field to be proportional to identity with a single scalar $\hat{\phi} = \phi \mathbb{1}$.

When both layers are described by the same parameter $\zeta$ (see Fig.\ref{fig:phi-Ds-minimal}{\bf a}), the stiffness has a significant enhancement coming from the geometric term $D_s^g$. 
The extra contribution is indeed proportional to the trace of the quantum metric in this particular case $D_s^g = (\phi^2/4A) \sum_{\bf k} g({\bf k})/E({\bf k})$ \cite{Hu2022}.
While the geometric contribution dominates at higher bias voltage $|V_b|$, it is important to note that a high $|V_b|$ results in higher densities of electron-hole pairs and may ultimately drive the system to the exciton Mott transition \cite{Mott1973, Brinkman_1973, Smith1986, Guerci2019}.

\section{Excitons with Asymmetric Bands}\label{app:asymmeyricBands}

\rev{
In this section, we discuss various aspects of particle-hole asymmetry in the geometric contribution to stiffness.

Let us begin by ignoring the effects of wavefunctions and recall the theory of excitons with asymmetric electron and hole masses.
This is a common} scenario in most semiconductors where the conduction and valence bands exhibit different dispersions, characterized by $m_e \neq m_h$. 
It leads to a slightly generalized form of Eq.~\eqref{eq:MFpb} with the mean-field Hamiltonian:
\begin{equation}
    \mathcal{H}_{\rm MF} = \sum\limits_{ {\bf k} } \psi^\dag_{{\bf k}} \Big[ \dfrac{{\bf k}^2}{2M} \mathbb{1} + \Big(\dfrac{{\bf k}^2}{2m} + V_b - \phi_z \Big) \tau_z + \phi \tau_x \Big] \psi^\phdag_{{\bf k}}.
\end{equation}
Here, $M^{-1}= (m_e-m_h)/2m_em_h$ and $m^{-1} = (m_e+m_h)/2m_em_h$ are effective masses. The inclusion of the identity term modifies the mean-field dispersion relation $E_\pm({\bf k}) = \xi_M({\bf k}) \pm E({\bf k})$, where $E({\bf k}) = \sqrt{\xi({\bf k})^2 + \phi^2}$ represents the previous quasiparticle dispersion and $\xi_M({\bf k}) = {\bf k}^2/2M$.
The self-consistent order parameter equation changes as well, mainly via the thermal factors, which were suppressed in Eq.~\eqref{eq:gap-eq-pb},
\begin{equation}
    \dfrac{1}{V} = \dfrac{1}{N}\sum\limits_{{\bf k}} \dfrac{ f[E_-({\bf k})]- f[E_+({\bf k})] }{ 2 E({\bf k}) }.
\end{equation}
It is easy to check that the equation reduces to Eq.~\eqref{eq:gap-eq-pb} if $E_+({\bf k}) = - E_-({\bf k})=E({\bf k})$. 
The effects stemming from the asymmetry can be visualized as a momentum-dependent Zeeman term in analogy to the one encountered in BCS theory.
\rev{It is well known that large anisotropy can act against exciton binding \cite{Chandrasekhar1962,Clogston1962} and destroy the condensate.
Phase stiffness is affected as well since
it is a convolution of $\phi$ with the counterflow current operator $\hat{j}_{\rm CF}$. 
If $\phi$ decreases, $D_s$ will decrease as well. }

\begin{figure}
    \centering
    \includegraphics[width=0.95\textwidth]{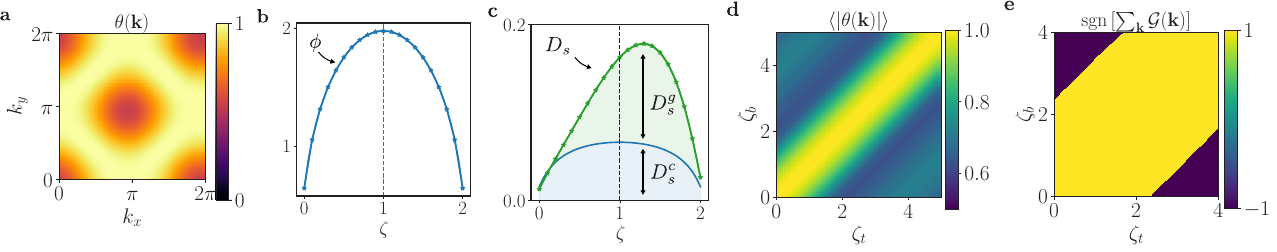}
    \caption{Asymmetric excitons in the minimal model with tunable quantum metric. {\bf a} The overlap function between two bands defined by $\zeta_t = 1$, $\zeta_b = \zeta = 2$. 
    The gap equation ({\bf b}) and stiffness ({\bf c}) are both modified as the asymmetry in wavefunction is tuned by $\zeta$ (with $V_b=0.1t$). 
    Notice that the conventional contribution is symmetric around $\zeta_t = \zeta_b =1$. {\bf d} Overlap function averaged over the BZ. Departure from unity results in a smaller mean-field solution for $\phi$. The isotropic case $\zeta_t = \zeta_b$ results in the strongest exciton. 
    {\bf e} Sign of the geometric contribution in the minimal model. There are parameter regions where the contribution is negative. 
    Intuitively, quantum geometry can be detrimental to stiffness if the bands are very different.}
    \label{fig:zetaTB}
\end{figure}

Next, we consider the more subtle scenario of symmetric energies but asymmetric wavefunctions.
We again utilize the minimal model presented in the previous section, eq.~\eqref{eq:def:zeta-ham}, with $\zeta_t = 1$ for the top layer and $\zeta_b = \zeta$ in the bottom layer. 
This configuration results in a particle-hole symmetric spectrum but with distinct electron and hole wavefunctions.
The overlap $\theta({\bf k}) = \cos[(\zeta-1) p_{\bf k}/2]$ exhibits significant variation across the Brillouin zone (see Fig.~\ref{fig:zetaTB}{\bf a}). 
This variation plays a crucial role in the quasiparticle energy $E({\bf k}) = \sqrt{\xi({\bf k})^2 + \phi^2 |\theta({\bf k})|^2}$, which, in turn, affects the gap equation $1/V = (1/N)\sum_{\bf k}1/[E({\bf k})]$. 
The reduction in $|\theta({\bf k})|$ from unity leads to a decrease in $\phi$ compared to the particle-hole symmetric case where $\theta({\bf k})=1$ (see Fig.~\ref{fig:zetaTB}{\bf b} and Fig.~\ref{fig:zetaTB}{\bf d}).
The stiffness is also affected, as shown in Fig.~\ref{fig:zetaTB}{\bf c}. 
Two factors contribute to this alteration: pre-factor $\phi$ and $\mathcal{G}({\bf k})$ which explicitly depends on $\zeta$ through $\theta({\bf k})$, $\mathcal{P}({\bf k})$, and $\mathcal{D}({\bf k})$.
Although it may be model specific, we find that the first contribution is symmetric around $\zeta=1$ while the geometric contibution is skewed towards higher $\zeta$ (see Fig.~\ref{fig:zetaTB}{\bf c}).
Additionally, it is important to highlight that the geometric contribution is not necessarily positive definite, see Fig.~\ref{fig:zetaTB}{\bf e}. 
This stands in contrast to the geometric contributions reported in superfluid stiffness \cite{Torma2021} and electron-phonon coupling \cite{yu2023}.

\rev{
Motivated by these two simple scenarios, we can deduce what happens in a realistic situation with both asymmetric dispersions and wavefunctions.
The dispersion asymmetry results in redefinition of effected masses $m$ and $M$.
On top of that, the wavefunction asymmetry modifies both $\phi$ and $\mathcal{G}({\bf k})$.
Both these effects are at play simultaneously, the details of which depend on the model and we are left limited to numerical results. 
}

\section{Continuum model for exciton condensation in AA-stacked TMDs}\label{app:ContinuumModel}

To start with we consider spinless electrons described by the effective Hamiltonian: 
\begin{equation}
       H=\sum^\Lambda_{ {\bk} ,a}{ \psi}^\dagger_{ {\bk}, a}
    \begin{bmatrix}
        H_{t,a}({\bk})-\mu+E_z& 0 \\ 0 & H_{b,a}({\bk})-\mu-E_z
    \end{bmatrix} { \psi}_{{\bk}, a}
    +\frac{1}{2A}\sum^\Lambda_{{\bq}}\sum_{\nu,\nu^\prime=t,b}V^{\nu\nu^\prime}_{\bq}n_{{\bq},\nu}n_{-{\bq},\nu^\prime},
\end{equation}
where $\psi_{k,a}$ is a 4-dim spinor in the sublattice and layer indeces, $\psi_{k,a}=[\psi_{k,a+t},\psi_{k,a-t},\psi_{k,a+b},\psi_{k,a-b}]$, $K$ and $K'$ refers to the two different valleys, $\Lambda$ is the UV cutoff and $E_z$ the electric displacement field controlling the energy offset between the two layers. The interaction is obtained by taking the Fourier transform of $V^{tt}_r=V^{bb}_r= e^2/(4\pi\epsilon_0\epsilon r)$ and $V^{tb}_r=V^{bt}_r= e^2/(4\pi\epsilon_0\epsilon  \sqrt{r^2+d^2})$ with $d$ interlayer distance. From
\begin{equation}
    \int\frac{d^2{\bf q}}{2\pi} \frac{e^{-dq+i{\bf q}\cdot{\bf r}}}{q} =\oint_{|z|=1}\frac{2dz}{2\pi r}\frac{1}{z^2+1-2dz/(ir)}=\frac{1}{\sqrt{r^2+d^2}},
\end{equation}
we readily find: 
\begin{equation}
\label{interaction_matrix}
    V^{tt}_{\bq}=V^{bb}_{\bq}=V_{\bq}=\frac{ e^2}{2\epsilon_0\epsilon q}, \quad V^{tb}_{\bq}=e^{-qd}V_{\bq}.
\end{equation}
Accounting for the presence of the metallic gates at distance $\xi$ from the sample we replace $1/q\to \tanh(q\xi/2)/q$, which regularizes the singularity at $q=0$. We use $\sigma$ for $d_{z^2}/d_{{x^2-y^2}}\pm id_{{xy}}$ at valley $K/K'$ while $\tau$ for layer. The chemical potential $\mu$ enforces the charge neutrality condition: 
\begin{eqnarray}
  \frac{1}{N}  \sum^\Lambda_{\bk}\langle \psi^\dagger_{{\bk},a}\psi_{{\bk},a}\rangle - n_{\rm CNP}= 0. 
\end{eqnarray}

\subsubsection{Model for TMDs}

 We assume that due to the strong Ising spin orbit coupling the conduction (CB) and valence band (VB) of each layer around $K$ and $K'$ points are spin polarized. This assumption is realized under the assumption of large Ising spin-orbit coupling $\tau\Delta_{vb}s^z$ and $\tau\Delta_{cb}s^z$ with $\tau=\pm$ for the two valleys. For MX$_2$ the SOC gap is $2\Delta_{vb}=148,186,219$meV with X=$\{\text{S}_2,\text{Se}_2,\text{Te}_2\}$. For WX$_2$ the gap is larger due to the larger SOC $2\Delta_{vb}=429,466,484$meV for VB and $2\Delta_{cb}=-32,-37,-53$meV for CB with X=$\{\text{S}_2,\text{Se}_2,\text{Te}_2\}$. In the conduction band the SOC is small and corrections from the nearby conduction band might play an important role. 
Keeping terms up to the second order in momentum the $k.p$ Hamiltonian reads:
\begin{equation}
\label{app:kp_Hamiltonian}
       H_{\nu,K}({\bk})=\begin{bmatrix}
        \Delta_\nu+\alpha_\nu k^2 & v_\nu k_- \\ v_\nu k_+ & -m_\nu-\gamma_\nu k^2\end{bmatrix}, \quad  H_{t/b,K'}({\bk})= H^*_{t/b,K}(-{\bk}),
\end{equation}
where the latter relation enforces the time reversal symmetry. The typical values of the parameters of the model are given in Table~\ref{tab:continuum_parameters}. Different configurations are characterized by a typical value of the ratio $v/m\approx 2\text{\AA}$. Finally, in the main text we compare results obtained with gapped Dirac cones and parabolic bands. The latter is obtained expanding the Dirac dispersion around $K$ and $K'$, $\pm\sqrt{\Delta^2+v^2 k^2}\simeq \pm\Delta \pm k^2/2 m$ with $m = \Delta /v^2$ and gap $2\Delta$.

\begin{table}[]
\centering
\begin{tabular}{|c||c|c|c|c|c|c|}
\hline
 & MoS$_2$ & MoSe$_2$ &  MoTe$_2$ & WS$_2$ & WSe$_2$ &  WTe$_2$   \\
\hline\hline
$2m$[eV]&   
 2.76 & 2.33 &  1.82&  2.88 & 2.42   &  1.82  \\
\hline
$v$[eV\,\AA] & 2.22 & 2.20 & 2.16 & 2.59 & 2.60 & 2.79  \\
\hline 
$\alpha$[eV\,\AA$^2$] &  0.52 &     -0.54 &  -1.66 & 2.03  &  1.08 & 0.17  \\
\hline
$\gamma$[eV\,\AA$^2$] & -6.65 & -6.20  & -5.78  &  -7.96  & -7.58  &  -7.31    \\
\hline
\end{tabular}
\caption{Relevant parameters of the $k.p$ Hamiltonian for different TMDs~\cite{Korm_nyos_2015} computed performing GW calculations~\cite{PhysRevB.85.205302,Liang_2013}.
} 
\label{tab:continuum_parameters}
\end{table}

\subsubsection{Symmetries}

The model is invariant under the transformation: 
\begin{equation}
    \psi_{{\bk},at}\to e^{i\phi_t}\psi_{{\bk},at},\quad \psi_{{\bk},at}\to e^{i\phi_b}\psi_{{\bk},at},
\end{equation}
where $\phi_b+\phi_t$ is the total phase associate to U(1)$_{\rm charge}$ while the relative phase $\phi_t-\phi_b$ is related to the U(1)$_{\rm layer}$ symmetry. 
The latter is spontaneously broken in the excitonic state which is characterized by interlayer coherence. In addition, we observe that the model in Eq.~\ref{app:kp_Hamiltonian} breaks the particle-hole symmetry for $\alpha\neq\gamma$. 
The low-energy model is also invariant under the continuous space rotational rotational symmetry ${\bk}\to R_\theta {\bk}$ and $U_\theta=\text{diag}\left(e^{-i\theta/2},e^{i\theta/2}\right)$ which acts on the sublattice degree of freedom:
\begin{equation}
    U^\dagger_\theta H_{\nu,K/K'}(R_\theta  {\bk} ) U_{\theta}= H_{\nu,K/K'}( {\bk} ).
\end{equation}
Including higher order terms in the $k.p$ expansion lower the aforementioned symmetry to $C_{3z}$. Finally,  $\mathcal M$ acting as $(x,y)\to(-x,y)$ and $\uparrow\leftrightarrow\downarrow$ is a symmetry of the model.

\subsubsection{Diagonalizing the single-particle Hamiltonian}

The eigenvalues of the non-interacting Hamiltonian are given by $E_{\pm,t/b}(k)=\pm\epsilon_{t/b}(k)$, $\epsilon_{t/b}(k)=\sqrt{v^2_{t/b}k^2+\Delta_{t/b}^2}$ while the matrix of eigenstates is:
\begin{equation}
    U({\bk})
    =\frac{1}{\sqrt{2}}\begin{bmatrix} \cos\frac{\varphi_k}{2}+\sin\frac{\varphi_k}{2} &  \cos\frac{\varphi_k}{2}-\sin\frac{\varphi_k}{2}   \\
   e^{i\phi_k} \left(\cos\frac{\varphi_k}{2}-\sin\frac{\varphi_k}{2}\right) & -e^{i\phi_k} \left(\cos\frac{\varphi_k}{2}+\sin\frac{\varphi_k}{2}\right) \end{bmatrix}
   ,
\end{equation}
where $\tan\varphi_k=\Delta/(vk)$ and we dropped the layer and valley labels. Applying the unitary transformation: 
\begin{equation}
\label{eq:diag_kin}
    U({\bk}) =\begin{bmatrix}
        U_t({\bk}) & 0 \\
        0 & U_b({\bk}) 
    \end{bmatrix},
\end{equation}
the kinetic part becomes: 
\begin{equation}
    H_{\rm kin}=\sum_{a=K,K'}\sum^\Lambda_{ \bk}\Psi^\dagger_{ {\bk}, a}
    \begin{bmatrix}
        \varepsilon_t(k)\sigma^z+(V_b-\mu)\sigma^0 & 0 \\ 0 & \varepsilon_b(k)\sigma^z+(-V_b-\mu)\sigma^0
    \end{bmatrix}\Psi_{{\bk}, a},
\end{equation}
where $\Psi_{{\bk},a \nu}=U^\dagger_{a \nu}(k)\psi_{{\bk},a \nu}$ with $a=K/K'$ and $\nu=t,b$. We now look at the interaction projected in the basis of the eigenstates of the kinetic term. The density operator for the layer $\nu=t/b$ reads: 
\begin{equation}
\rho^\nu_{\bq}=\sum_{\bk}\psi^\dagger_{{\bk} \nu}\psi_{{\bk}+{\bq} \nu}=\sum_{\bk}\sum_{\alpha\alpha'} \left[F^\nu_{{\bk},{\bq}}\right]_{\alpha\alpha'} \Psi^\dagger_{{\bk} \nu\alpha}\Psi_{{\bk}+{\bq} \nu\alpha'},
\end{equation}
where we drop for the moment the valley index, and the form factor is obtained projecting the density operator in the Bloch basis
\begin{equation}
   F^\nu_{{\bk},{\bk}} = U^\dagger_\nu({\bk}) U_\nu({\bk}+{\bq}), \quad \left[F^\nu_{{\bk},{\bq}}\right]_{\alpha\alpha'} =\braket{u^\nu_{{\bk},\alpha}}{u^\nu_{{\bk}+{\bq},\alpha'}}. 
\end{equation}
The orthogonality condition implies that $\left[F^\nu_{{\bk},0}\right]_{\alpha\alpha'}=\delta_{\alpha\alpha'}$. 
As a result the interaction takes the form: 
\begin{equation}
    H_{\rm int} = \frac{1}{2A}\sum^\Lambda_{{\bq}{\bk}{\bk}'}\sum_{\nu\nu^\prime}\sum_{aa'}\left[F^{\nu,a}_{{\bk},{\bq}}\right]_{\alpha\alpha'}V^{\nu\nu^\prime}_{\bq} \left[F^{\nu^\prime,a'\dagger}_{{\bk}',{\bq}}\right]_{\gamma\gamma'} \Psi^\dagger_{{\bk},a\nu\alpha}\Psi_{{\bk}+{\bq},a\nu\alpha'}\Psi^\dagger_{{\bk}'+{\bq},a'\nu^\prime\gamma}\Psi_{{\bk}',a'\nu^\prime\gamma'},
\end{equation}
At mean-field level we decouple the interaction in the Hartree and Fock (exchange) terms that are derived in the following sections. 

\subsection{Hartree term}
\label{app:Hartree_term}

The Hartree term corresponds to: 
\begin{equation}
H_{H}=\sum^{\Lambda}_{\mathbf k}\sum_{a=K,K'}\sum_{\nu=t,b} \delta\mu_\nu\,\Psi^\dagger_{{\mathbf k},a\nu}\Psi_{{\mathbf k},a\nu}-\frac{1}{2A}\sum_{\nu\nu^\prime}V^{\nu\nu^\prime}_0\sum^{\Lambda}_{{\mathbf k}{\mathbf k}'}\sum_{aa'=K,K'}\delta \rho^{a\nu}({\mathbf k}) \delta \rho^{a'\nu^\prime}({\mathbf k}'),
\end{equation}
where we have introduced the charge density measured with respect to charge neutrality: 
\begin{equation}
    \delta\rho^{a\nu}({\mathbf k})=\sum_{\alpha=\pm}\langle\Psi^\dagger_{{\mathbf k},a\nu\alpha}\Psi_{{\mathbf k},a\nu\alpha}\rangle-N\delta_{{\mathbf k},0}.
\end{equation}
the contribution $\delta\mu_\nu$ is given by: 
\begin{equation}
    \delta\mu_\nu = \frac{1}{N}\sum^\Lambda_k\sum_{a=K,K'}\sum_{\alpha=\pm}\sum_{\nu^\prime}\delta \rho^{a\nu^\prime}_{\alpha\alpha}({\mathbf k})V^{\nu^\prime\nu}_0.
\end{equation}
Before moving on we observe that 
\begin{equation}
    \frac{1}{N}\sum^\Lambda_k\sum_{a=K,K'}\sum_{\alpha=\pm}\delta \rho^{a}_{\alpha\alpha}({\mathbf k})=n_\nu-1=\text{sign}(\nu) \phi_z,
\end{equation}
where $\phi_z$ measure the charge density transferred between the layers $\phi_z=(n_t-n_b)/2=n_t-1=1-n_b$ since at charge neutrality $n_t+n_b=2$ with this definition $\text{sign}(t)=1$ and $\text{sign}(b)=-1$. Thus, we have: 
\begin{equation}
    \delta\mu_t=\phi_z(V^{tt}_0-V^{bt}_0),\quad \delta\mu_b=\phi_z(V^{tb}_0-V^{bb}_0). 
\end{equation}
From Eq.~\eqref{interaction_matrix} we find: 
\begin{equation}
    \delta\mu_t=\delta \mu ,\quad \delta\mu_b=-\delta\mu,\quad \delta\mu = \phi_z(V^{tt}_0-V^{bt}_0),\quad \phi_z=(n_t-n_b)/2,
\end{equation}
which corresponds to electrostatic repulsion. We notice that $V^{tt}_0=V^{tb}_0$ and $\delta\mu=0$.

\subsection{Fock term}
\label{app:fock_term}

\begin{figure}
    \centering
    \includegraphics[width=0.6\textwidth]{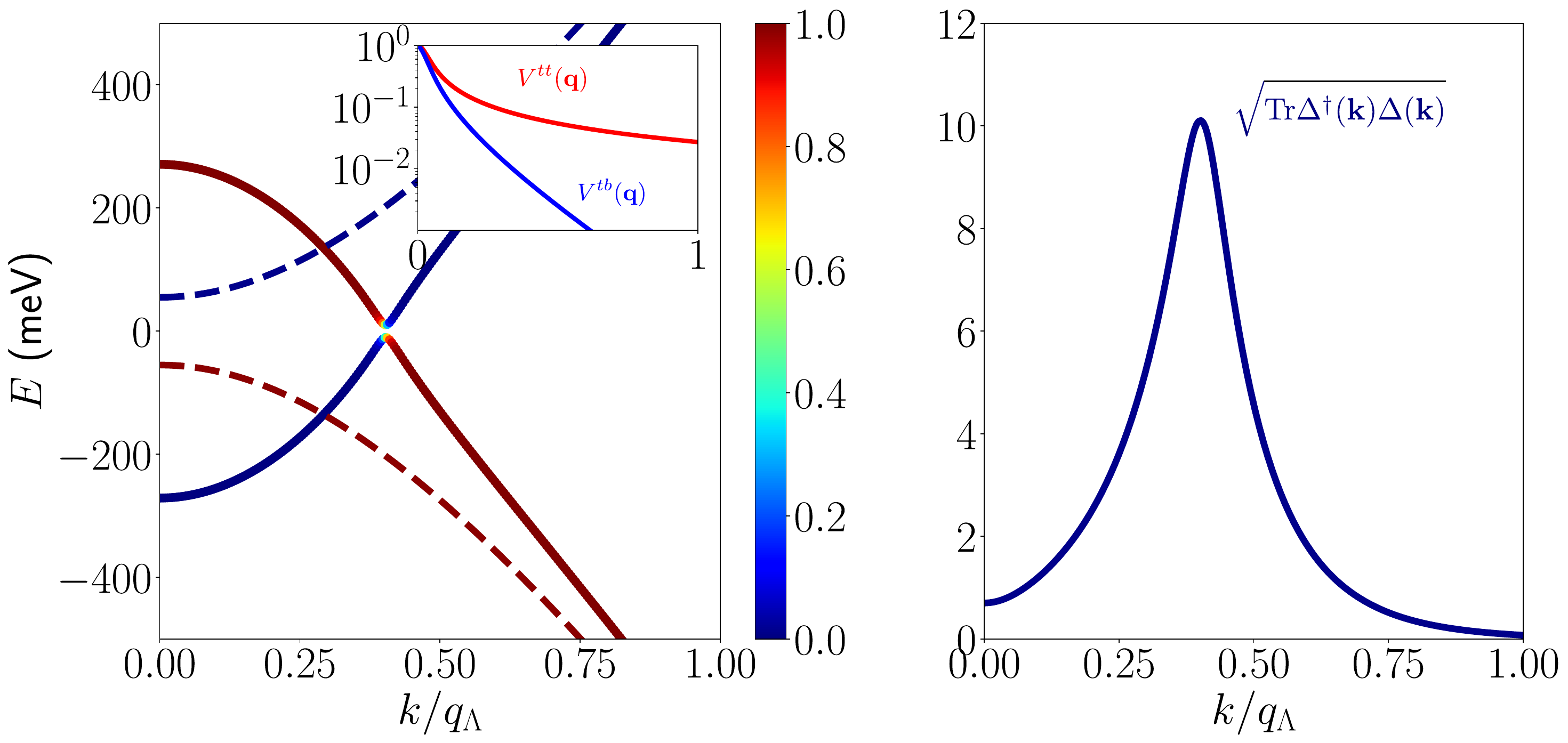}
    \caption{ a) Non-interacting band structure (dashed) and interacting excitonic energy (solid). The color denotes the layer polarization showing the mixing between the different bands in the excitonic state.
    The inset shows the intralayer and interlayer Coulomb interaction as a function of $q/q_\Lambda$ in unit of $V_{\bq=0}$. Notice that $V^{tb}_{\bf q}$ decays exponentially with $q$. b) Gap function $\sqrt{\Tr \Delta^\dagger({\bk})\Delta({\bk})}$ denoting interlayer phase coherence. Calculations are performed for $\Delta E=(\Delta_t+\Delta_b-2V_b)/2=55$meV where the single particle spectrum is gapped, $v_t=v_b=2.19$eV{\AA}, $\Delta_t=\Delta_b=0.9$eV, we set the screening length $\xi=12$nm and $d=$1nm. We fix the UV cutoff $q_\Lambda a=2\pi/3$ corresponding to half-distance between the two corners $K$ and $K'$ of the TMD Brillouin zone. These parameters are the same ad those employed for the results on homobilayer reported in the maintext.
    }
    \label{fig:continuum_bands}
\end{figure}

The exchange term is given by: 
\begin{equation}
    H_X=-\sum^\Lambda_k\sum_{a=K,K'}\sum_{\nu\nu^\prime}\Psi^\dagger_{{\mathbf k},a\nu\alpha}\Sigma^{\nu\nu^\prime,a}_{F,\alpha\beta}({\mathbf k})\Psi_{{\mathbf k},a\nu^\prime\beta},
\end{equation}
where we have introduced the quantity: 
\begin{equation}
\label{fock_selfenergy}
    \Sigma^{\nu\nu^\prime,a}_{X,\alpha\beta}({\mathbf k})=\frac{1}{A}\sum_{{\mathbf k}'} V^{\nu\nu^\prime}_{{\mathbf k}'-{\mathbf k}}\sum_{\alpha'\beta'} \left[F^{\nu^\prime,a\dagger}_{{\mathbf k},{\mathbf k}'-{\mathbf k}}\right]_{\beta'\beta}\left[F^{\nu,a}_{{\mathbf k},{\mathbf k}'-{\mathbf k}}\right]_{\alpha\alpha'} \delta \rho^{\nu^\prime\nu,a}_{\beta'\alpha'}({\mathbf k}'),
\end{equation}
where we have introduced the quantity: 
\begin{equation}
    \delta \rho^{\nu^\prime\nu,a}_{\beta'\alpha'}({\mathbf k})=\langle\Psi^\dagger_{{\mathbf k},a\nu^\prime\beta'}\Psi_{{\mathbf k},a\nu\alpha'}\rangle-\delta_{\nu,\nu^\prime}\delta_{\alpha'\beta'}\delta_{\alpha'-}.
\end{equation}
$\Sigma({\mathbf k})$ is a 4-dim matrix in the layer and band basis. It includes an intralayer terms which renormalize the electronic band structure and an interlayer term giving rise to interlayer phase coherence and exciton condensation. The Hartree-Fock Hamiltonian takes the form: 
\begin{equation}
\label{HF_continuum_model}
    H_{\rm HF}=\sum^\Lambda_k\Psi^\dagger_{ {\mathbf k}}
 \left[\hat H_{\rm kin}({\mathbf k})-\hat \Sigma({\mathbf k})\right]\Psi_{{\mathbf k}}.
\end{equation}
The self-consistency equation~\ref{fock_selfenergy} for the self-energy is solved at every ${\mathbf k}$-point in momentum space employing an iterative approach.
Fig.~\ref{fig:continuum_bands} shows the typical band structure and the interlayer gap function, defined as $\sqrt{\Tr \Delta^\dagger({{\mathbf k}})\Delta({{\mathbf k}})}$ with $\hat \Delta({\mathbf k})$ interlayer component of the self-energy $\hat \Delta({\mathbf k})=\text{Tr}\left[\tau^+\hat \Sigma({\mathbf k})\right]$. Finally, we remark that the excitonic order parameter $\phi$ has been computed as: 
\begin{equation}
    \phi_{x}=\frac{T}{A}\sum_{{\mathbf k}}\sum_{i\omega} \text{Tr}\left[\tau^x\otimes\sigma^a \hat G({\mathbf k},i\omega)\right],
\end{equation}
in the gauge where the interlayer coherence is broken along $\tau^x$ and the Green's function $\hat G$ is obtained including all the bands of the model from Eq.~\eqref{HF_continuum_model}.

\end{appendix}

\end{document}